\documentclass[12pt]{article}

\usepackage{amsmath,amssymb,amsfonts}
\usepackage{psfrag}
\usepackage{color}
\definecolor{darkblue}{rgb}{0.1,0.1,.7}
\usepackage[colorlinks, linkcolor=darkblue, citecolor=darkblue, urlcolor=darkblue, linktocpage]{hyperref} 
\usepackage[square, comma, sort&compress,numbers]{natbib}
\usepackage[]{amsmath}
\usepackage[]{graphicx}
\usepackage[]{latexsym}
\usepackage{geometry}
\usepackage{amscd}
\usepackage[all,cmtip]{xy}
\usepackage{mathrsfs}
\usepackage[margin=10pt,font=small,labelfont=bf]{caption}
\usepackage{simplewick}
\usepackage{bbold}
\usepackage{changepage}
\usepackage{dsfont}
\usepackage{bm}
\usepackage{float}

\numberwithin{equation}{section}

\newcommand{\be}{\begin{eqnarray}}
\newcommand{\ee}{\end{eqnarray}}
\newcommand{\bea}{\begin{eqnarray}}
\newcommand{\eea}{\end{eqnarray}}

\newcommand   \SU    {\mathrm{SU}}
\newcommand \U {\mathrm{U}}
\newcommand   \SO    {\mathrm{SO}}

\newcommand   \Tr    {\mathrm{Tr}}

\renewcommand \a  {\alpha}
\renewcommand \b  {\beta}

\newcommand   \f  {\phi}

\newcommand   \s  {\sigma}

\def\beq{\begin{equation}} 
\def\eeq{\end{equation}} 
\def\<{\langle}
\def\>{\rangle}

\def\nn{\nonumber} 
 
\def\cO {{\cal O}}

\begin{document}

\vspace*{-.6in} \thispagestyle{empty}
\begin{flushright}
\end{flushright}
\vspace{.2in} {\Large
\begin{center}
{\bf Conformal Collider Physics from the \\Lightcone Bootstrap\vspace{.1in}}
\end{center}
}
\vspace{.2in}
\begin{center}
{\bf 
Daliang Li$^{a,b}$, David Meltzer$^a$, David Poland$^{a,c}$} 
\\
\vspace{.2in} 
$^a$ {\it  Department of Physics, Yale University, New Haven, CT 06511}\\
$^b$ {\it  Department of Physics and Astronomy, Johns Hopkins University, Baltimore, MD 21218}\\
$^c$ {\it School of Natural Sciences, Institute for Advanced Study, Princeton, NJ 08540}
\end{center}

\vspace{.2in}

\begin{abstract}
We analytically study the lightcone limit of the conformal bootstrap equations for 4-point functions containing global symmetry currents and the stress tensor in 3d CFTs. We show that the contribution of the stress tensor to the anomalous dimensions of large spin double-twist states is negative if and only if the conformal collider physics bounds are satisfied. In the context of AdS/CFT these results indicate a relation between the attractiveness of AdS gravity and positivity of the CFT energy flux. We also study the contribution of non-Abelian conserved currents to the anomalous dimensions of double-twist operators, corresponding to the gauge binding energy of 2-particle states in AdS. We show that the representation of the double-twist state determines the sign of the gauge binding energy if and only if the coefficients appearing in the current 3-point function satisfies a similar bound, which is equivalent to an upper bound on the charge flux asymmetry of the CFT. 
\end{abstract}

\newpage

\tableofcontents

\newpage

\section{Introduction}
\label{sec:intro}

The conformal bootstrap program \cite{Ferrara:1973yt, Polyakov:1974gs, Mack:1975je} has seen a marked revival in recent years for CFTs in $d>2$ spacetime dimensions. Using only unitarity and associativity of the operator product expansion (OPE), it was found in \cite{Rattazzi:2008pe} that one could derive numerical bounds on the spectrum of an arbitrary CFT by studying the four point function of identical scalars. The numerical work has been extended greatly to include global symmetries \cite{Vichi:2011ux,Nakayama:2014yia,Bae:2014hia,Rattazzi:2010yc,Kos:2013tga,Kos:2015mba,Nakayama:2014sba,Chester:2014gqa,Nakayama:2014lva,Chester:2015qca}, supersymmetry \cite{Alday:2013opa,Beem:2013qxa,Beem:2014zpa,Beem:2014kka,Beem:2015aoa,Chester:2014fya,Khandker:2014mpa,Beem:2013sza,Chester:2014mea,Fortin:2011nq,Fitzpatrick:2014oza,Beem:2014rza,Berkooz:2014yda,Poland:2010wg,Bobev:2015vsa,Bobev:2015jxa,Poland:2015mta,Lemos:2015awa,Lin:2015wcg}, and correlation functions of non-identical scalars \cite{Kos:2014bka}. In particular, there has been remarkable progress in numerically solving the 3d Ising \cite{Kos:2014bka,El-Showk:2014dwa,ElShowk:2012ht,Simmons-Duffin:2015qma} and critical O(N) models \cite{Kos:2013tga,Kos:2015mba}. Furthermore, it was found that there do exist limits where the conformal bootstrap equations can be solved analytically. The pertinent example for this paper is the lightcone limit, first studied in \cite{Fitzpatrick:2012yx,Komargodski:2012ek} and extended in \cite{Fitzpatrick:2014vua,Kaviraj:2015cxa,Alday:2015ota,Vos:2014pqa,Alday:2015eya,Kaviraj:2015xsa,Fitzpatrick:2015qma,Li:2015rfa,Alday:2015ewa}. In this limit the expansion parameter is the twist of the exchanged operator, as opposed to its dimension. In a unitary, interacting CFT in $d>2$ dimensions there exist a finite number of operators with very low twist, namely the identity operator, conserved operators, and possibly some low dimension scalars. This fact is crucial in solving the lightcone bootstrap equations.

With the exception of recent work on the four point function of 3d fermions \cite{Iliesiu:2015qra}, the bootstrap equations have primarily been studied for four external scalars. In this work we will focus instead on four point functions of 3d CFTs that include external conserved spin 1 and spin 2 operators, i.e. conserved currents $J_{\mu}$ and the stress-energy tensor $T_{\mu\nu}$. The restriction to 3d is technical, as in this dimension all the relevant conformal blocks are known \cite{Costa:2011dw}. We will study these equations analytically in the lightcone limit and solve for the anomalous dimensions of a wide variety of double-twist states. To be specific, we will study the correlation functions $\<JJ\f\f\>$, $\<JJJJ\>$, and $\<TT\f\f\>$ where $J$ is a conserved current for a global $\U(1)$ or $\SU(N)$ symmetry, $T$ is the stress energy tensor, and $\f$ is a scalar of arbitrary dimension. The s-channel conformal block expansion is dominated by the contribution from low twist operators such as the identity, the conserved currents $J$, the stress tensor $T$, and possibly some scalars with small dimensions. As in \cite{Fitzpatrick:2012yx,Komargodski:2012ek}, we show that large spin double-twist operators must exist to satisfy the crossing equations. Their anomalous dimensions are determined by the OPE coefficients of the lower twist operators exchanged in the s-channel. In an AdS dual description, the contributions to the anomalous dimensions from $J$ and $T$ correspond to the binding energies of well separated 2-particle states arising from gauge and gravitational interactions. 

An important feature of our results will be that the contributions of $J$ or $T$ exchange to the anomalous dimensions of double-twist states flip signs if the relevant s-channel OPE coefficients do not lie between the free boson and free fermion values. For the exchange of $T$, this means requiring that gravity in the bulk be attractive yields the Hofman-Maldacena conformal collider physics bounds~\cite{Hofman:2008ar} on coefficients appearing in $\<JJT\>$ and $\<TTT\>$ (extended to general dimensions in~\cite{Buchel:2009sk,Chowdhury:2012km}), which were originally discovered by requiring that the integrated energy flux produced by a localized perturbation in Minkowski space is positive.

Unlike the energy flux, it is not obvious if the integrated charge flux should have a definite sign. Consequently it is not clear if analogous bounds on the current three point function coefficients should hold. However, we will see that when these coefficients do not lie between the free field theory values, the contributions of $J$ to the anomalous dimensions would have counter-intuitive signs, motivating us to speculate that analogous bounds on the coefficients appearing in $\<JJJ\>$ might hold. A corollary of our results is that when these OPE coefficients saturate their free field theory values, some of the anomalous dimension asymptotics vanish, possibly indicating that subsectors or the entirety of the theory are free \cite{Zhiboedov:2013opa}.

The organization of this paper is as follows. In Section \ref{sec:correlation} we review the construction of correlation functions for operators with spin using an embedding formalism. By constructing a differential representation of the 3-point functions we can calculate the relevant conformal blocks. In Section \ref{sec:bootstrap} we review the lightcone bootstrap for four external scalars and generalize it to include external operators with spin. In Section \ref{sec:JJOO} we consider $\<JJ\f\f\>$ with $J$ being either a $\U(1)$ or a $\SU(N)$ conserved current. We first solve the crossing equation at leading order in the lightcone limit, where large spin double-twist operators must contribute in order to reproduce the identity contribution. We then solve the equation at the first subleading order, where the exchange of a conserved current and stress energy tensor in the s-channel is reproduced by the anomalous dimensions of the aforementioned double-twist operators. This procedure is repeated in Sections \ref{sec:JJJJ} and \ref{sec:TTOO} for $\<JJJJ\>$ and $\<TT\f\f\>$ respectively. We also generalize a sufficient condition for the existence of higher spin symmetry in the limit of large global symmetry groups discovered in \cite{Li:2015rfa}. In Section \ref{sec:SCFT} we discuss some applications of our results to superconformal field theories (SCFTs) and in Section \ref{sec:discussion} we discuss some implications of our results and possible future work. In the appendices we collect technical details referenced in the paper.

\section{Correlation Functions}
\label{sec:correlation}

\subsection{The Embedding Formalism}

We will first review the embedding space formalism for CFTs developed in~\cite{Costa:2011dw,Costa:2011mg}. The idea, first noted by Dirac \cite{Dirac:1936}, is that the conformal group in $d$ Euclidean dimensions, $\SO(d+1,1)$, can be realized linearly in an embedding space $\mathbb{M}^{d+2}$ as the group of isometries. The constraints on correlation functions of primary operators simplifies to the constraints of Lorentz symmetry once we lift the fields to the embedding space. In this paper we will be interested in CFTs living in a $d$-dimensional Minkowski spacetime with conformal group $\SO(d,2)$, but we can always Wick rotate between the two pictures. 

The lift of $\mathbb{R}^{d}$ to $\mathbb{M}^{d+2}$ is accomplished by identifying points $x$ in $\mathbb{R}^{d}$ with null rays in $\mathbb{M}^{d+2}$ as
\begin{eqnarray}
P^{A}=\lambda(1,x^{2},x^{a}), \ \ \ \lambda \in \mathbb{R}, \ \ P^{A}\in\mathbb{M}^{d+2},
\end{eqnarray}
where $P^{A}$ is written in the lightcone basis
\bea
P^{A}=(P^{+},P^{-},P^{a}),
\eea
with the metric given by
\bea
P\cdot P \equiv \eta_{AB}P^{A}P^{B}=-P^{+}P^{-}+\delta_{ab}P^{a}P^{b}.
\eea

A linear $\SO(d+1,1)$ transformation maps null rays to null rays and therefore defines a transformation of the physical space onto itself. It can be shown that any $\SO(d+1,1)$ transformation of $\mathbb{M}$ induces a conformal transformation on $\mathbb{R}$ and that every conformal transformation can be obtained in this way.

We now need to give the correspondence between fields in the physical space and those in the embedding space. This was done using an index-free notation for symmetric traceless tensor fields in \cite{Costa:2011dw,Costa:2011mg} and has also been generalized to arbitrary tensor fields \cite{Costa:2014rya}, spinors in three \cite{Iliesiu:2015qra} and four dimensions \cite{Weinberg:2010fx,SimmonsDuffin:2012uy,Elkhidir:2014woa}, and various situations with supersymmetry, e.g.~\cite{Goldberger:2011yp, Siegel:2012di, Maio:2012tx, Kuzenko:2012tb, Goldberger:2012xb, Khandker:2012pa, Fitzpatrick:2014oza} and references therein. Three dimensions is special because the only irreducible tensor representations of $\SO$(3) are the totally symmetric and traceless representations. We only study 3d CFTs in this paper so we will restrict our review to these representations.

The mapping is as follows. Consider a field $F_{A_{1}...A_{\ell}}(P)$, a tensor of $\SO(d+1,1)$, with the following properties:

	1. Defined on the cone $P^{2}$=0 , \\ 
	\indent	2. Homogeneous of degree --$\Delta$: $F_{A_{1}...A_{\ell}}(\lambda P)=\lambda^{-\Delta}F_{A_{1}...A_{\ell}}(P), \ \ \lambda > 0$ ,  \\
	\indent	3. Symmetric and traceless , \\ 
	\indent	4. Transverse: ($P\cdot F$)$_{A_{2}...A_{\ell}}\equiv P^{A}F_{AA_{2}...A_{\ell}}=0 .$

Now we define the Poincar\'e section as
\bea
P_{x}^{A}=(1,x^{2},x^{a}), \ \ x\in\mathbb{R}^{d} .
\eea
Due to the homogeneity property, once $F$ is known on the Poincar\'e section it is known everywhere on the lightcone. The projection to the Poincar\'e section defines a symmetric traceless field on $\mathbb{R}^{d}$,
\bea
f_{a_{1}...a_{\ell}}(x)=\frac{\partial P^{A_{1}}}{\partial x^{a_{1}}} ... \frac{\partial P^{A_{\ell}}}{\partial x^{a_{\ell}}}F_{A_{1}...A_{\ell}}(P_{x}) .
\eea
To encode the symmetric traceless tensors in terms of a polynomial we introduce an auxiliary complex polarization vector $z^{a}$, contract it with the tensor field, and restrict to the submanifold defined by $z^{2}=0$,
\bea
f_{a_{1}...a_{\ell}} \ \ \text{symmetric and traceless} \ \leftrightarrow f(z)\big{|}_{z^{2}=0}.
\eea
We do not lose any information with this condition since our tensors will be traceless. Two polynomials that differ by terms that vanish when $z^{2}=0$ correspond to the same tensor. There is in fact a one to one correspondence between symmetric traceless tensors $f_{a_{1}...a_{\ell}}$ and polynomials $f(z)\big{|}_{z^{2}=0}$. The same idea is applied for tensors in the embedding space and we have that
\bea
F_{A_{1}...A_{\ell}}(P) \ \ \text{symmetric and traceless} \ \leftrightarrow F(P,Z)\big{|}_{Z^{2}=0,Z\cdot P=0} .
\eea
Once again we restrict to $Z^{2}=0$ since the tensor will be traceless and $Z\cdot P=0$ since it is transverse. That is, the polynomial is invariant under $Z\rightarrow Z + \lambda P$. Any polynomial that differs from $F(P,Z)$ by such terms corresponds to the same underlying tensor field. Defining $Z_{z,x}\equiv(0,2x\cdot z,z)$, the correspondence between the polynomials can be stated as
\bea
f(x,z)=F(P_{x},Z_{x,z}).
\eea

\subsection{3-point Functions}\label{subsec:3-pointFunctions}
In embedding space the classification of 3-point functions simplifies. Conformal symmetry fixes the basic building blocks for symmetric, traceless fields to be:
\bea
H_{ij}\equiv -2\big[(Z_{i}\cdot Z_{j})(P_{i}\cdot P_{j})-(Z_{i}\cdot P_{j})(Z_{j}\cdot P_{i})\big],  \\
V_{i,jk}\equiv \frac{(Z_{i}\cdot P_{j})(P_{i}\cdot P_{k})-(Z_{i}\cdot P_{k})(P_{i}\cdot P_{j})}{(P_{j}\cdot P_{k})}.
\eea
To simplify notation we define $P_{ij}=-2P_{i}\cdot P_{j}$. When projected to the Poincar\'e section we have $P_{ij}\rightarrow x_{ij}^{2}$ with $x_{ij}\equiv x_{i}-x_{j}$. The 3-point function can be written as
\bea\label{eq:3pfForm}
G_{\chi_{1},\chi_{2},\chi_{3}}(\{P_{i};Z_{i}\})=\frac{Q_{\chi_{1},\chi_{2},\chi_{3}}(\{P_{i},Z_{i}\})}{(P_{12})^{\frac{\s_{1}+\s_{2}-\s_{3}}{2}}(P_{23})^{\frac{\s_{2}+\s_{3}-\s_{1}}{2}}(P_{31})^{\frac{\s_{1}+\s_{3}-\s_{2}}{2}}},
\eea
where $\s_{i}=\Delta_{i}+\ell_{i}$. 
Defining
\bea
V_{1}\equiv V_{1,23}, \ \ \ V_{2}\equiv V_{2,31}, \ \ \ V_{3}\equiv V_{3,12},
\eea
then $Q$ can be written as a linear combination of structures of the form
\bea\label{eq:3pfStandardBasis}
\prod_{i}V_{i}^{m_{i}}\prod_{i<j}H_{ij}^{n_{ij}},
\eea
where the homogeneity properties of the operators imply
\bea
m_{i}+\sum_{j\neq i}n_{ij}=\ell_{i}.
\eea
In three dimensions some of these tensor structures are degenerate. In particular, 
\bea
(V_{1}H_{23}+V_{2}H_{13}+V_{3}H_{12}+2V_{1}V_{2}V_{3})^{2} = -2H_{12}H_{13}H_{23} + O(\{Z_{i}^{2},Z_{i}\cdot P_{i}\}).
\label{eqn:3D-deg}
\eea
We discuss these degeneracy conditions in more detail in Appendix \ref{App:DegeneracyEquations}.
 
There is an alternative way to represent the 3-point functions that will be useful for constructing conformal blocks. When the 3rd operator is symmetric and traceless, the spinning 3-point function can be expressed in terms of differential operators acting on an appropriate scalar 3-point function:
\be
\langle\phi_{1}^{\{a\}}(x_1)\phi_{2}^{\{b\}}(x_2)\mathcal{O}_{\{e\}}(x_3)\rangle=D^{a,b}_{x_{1},x_{2}}\langle\phi_{1}^{\prime}(x_1)\phi_{2}^{\prime}(x_2)\mathcal{O}_{\{e\}}(x_3)\rangle.
\ee
Explicit construction of these operators can be done in the embedding space, where they satisfy the consistency conditions $D \mathcal{O} (P_{i}^2, P_i\cdot Z_i, Z_{i}^2) = \mathcal{O} (P_{i}^2, P_i\cdot Z_i, Z_{i}^2)$. Such operators can be built out of the building blocks 
\bea \small
\begin{split}
& D_{11}\equiv (P_{1}\cdot P_{2})(Z_{1}\cdot\frac{\partial}{\partial P_{2}})-(Z_{1}\cdot P_{2})(P_{1}\cdot\frac{\partial}{\partial P_{2}})-(Z_{1}\cdot Z_{2})(P_{1}\cdot\frac{\partial}{\partial Z_{2}})+(P_{1}\cdot Z_{2})(Z_{1}\cdot\frac{\partial}{\partial Z_{2}}),
&\\&
D_{12}\equiv (P_{1}\cdot P_{2})(Z_{1}\cdot\frac{\partial}{\partial P_{1}})-(Z_{1}\cdot P_{2})(P_{1}\cdot\frac{\partial}{\partial P_{1}})+(Z_{1}\cdot P_{2})(Z_{1}\cdot\frac{\partial}{\partial Z_{1}}).
\end{split} 
\eea
There are two more operators $D_{22}$ and $D_{21}$ which can be found by permuting $1\leftrightarrow 2$. $D_{ij}$ acts to increase the spin at point $i$ by 1 and decreases the dimension at point $j$ by 1. A fifth operator is multiplication by $H_{12}$ which increases the spin and decreases the dimension at both points by one. The most general parity even spinning 3-point function can be constructed with the following basis\footnote{The operators can be reordered, keeping in mind two pairs do not commute: $[D_{11},D_{22}]\neq 0$ and $[D_{12},D_{21}]\neq 0$. All other differential operators commute with each other.}: 
\bea
&H_{12}^{n_{12}}D_{12}^{n_{10}}D_{21}^{n_{20}}D_{11}^{m_{1}}D_{22}^{m_{2}}\Sigma^{m_{1}+n_{20}+n_{12},m_{2}+n_{10}+n_{12}}\langle\phi_{1}(P_1)\phi_{2}(P_2)\mathcal{O}(P_3,Z_3)\rangle,
\eea 
where $m_{1}+n_{10}+n_{12}=\ell_{1}$ and $m_{2}+n_{20}+n_{12}=\ell_{2}$. The $\Sigma^{a,b}$ operators shift the dimensions by $\Delta_{1}\rightarrow\Delta_{1}+a$ and $\Delta_{2}\rightarrow\Delta_{2}+b$. We call this the differential basis. The transformation to the standard basis, (\ref{eq:3pfForm}) and (\ref{eq:3pfStandardBasis}), is computed in \cite{Costa:2011mg}.

In three dimensions we also have parity odd structures, which are given by the parity even tensor structures above multiplied by the epsilon tensor. In the standard basis we have the 3-point function structure
\bea
\epsilon_{ij}\equiv P_{ij}\epsilon(Z_{i},Z_{j},P_{1},P_{2},P_{3}),
\eea
where on the right hand side we have used the 5d epsilon tensor. We could also consider the structure formed with three $Z$ vectors and two $P$ vectors, but these can always be solved for in terms of $\epsilon_{ij}$. Similarly we can always solve for $\epsilon_{12}$ in terms of $\epsilon_{13}$ and $\epsilon_{23}$ \cite{Costa:2011dw,Costa:2011mg}. Therefore, we only need to use $\epsilon_{13}$ and $\epsilon_{23}$ to construct parity odd 3-point functions. Note that when multiplying by $\epsilon_{ij}$ the scaling dimensions must be shifted to preserve the desired scaling properties.

The corresponding parity odd differential operators are\footnote{There are two more differential operators, $\tilde{D}_{121}=\epsilon(Z_{1},Z_{2},P_{1},P_{2},\frac{\partial}{\partial P_{1}})$ and $\tilde{D}_{122}=\epsilon(Z_{1},Z_{2},P_{1},P_{2},\frac{\partial}{\partial P_{2}})$. For this paper, they can be ignored since their action on the scalar 3-point functions can be re-expressed in terms of the first two operators.}:
\be
&\tilde{D}_{1} \equiv \epsilon\bigg(Z_{1},P_{1},\frac{\partial}{\partial P_{1}},P_{2},\frac{\partial}{\partial P_{2}}\bigg)+\epsilon\bigg(Z_{1},P_{1},\frac{\partial}{\partial P_{1}},Z_{2},\frac{\partial}{\partial Z_{2}}\bigg),
\\
&\tilde{D}_{2} \equiv \epsilon\bigg(Z_{2},P_{2},\frac{\partial}{\partial P_{2}},P_{1},\frac{\partial}{\partial P_{1}}\bigg)+\epsilon\bigg(Z_{2},P_{2},\frac{\partial}{\partial P_{2}},Z_{1},\frac{\partial}{\partial Z_{1}}\bigg).
\ee
$\tilde{D}_i$ increases the spin at point $i$ by 1. To construct a parity odd 3-point function we act with a single odd differential operator, $\tilde{D}_{1}$ or $\tilde{D}_{2}$, and then the parity even operators. 

Finally, we will be interested in studying correlation functions involving conserved currents. As explained in  \cite{Costa:2011dw, Costa:2011mg}, requiring that a 3-point function be conserved at point $P_{i}$ is equivalent to requiring that $\partial_{P_{i}}\cdot D_{Z_{i}}$ vanish when acting on the embedding space correlation function, where
\bea
\partial_{P}\cdot D_{Z}\equiv\frac{\partial}{\partial P_{M}}\bigg[\bigg(\frac{d}{2}-1+Z\cdot\frac{\partial}{\partial Z}\bigg)\frac{\partial}{\partial Z^{M}}-\frac{1}{2}Z_{M}\frac{\partial^{2}}{\partial Z\cdot \partial Z}\bigg] .
\eea

\subsection{4-point Functions}

In this section we will review the structure of conformal blocks that appear in the 4-point functions of scalars as well as how to construct the conformal blocks for external operators with spin. First we start with four distinct scalars with dimensions $\Delta_{i}$, so the four point function is fixed by conformal invariance to be of the form
\bea
\<\f_{1}(x_{1})\f_{2}(x_{2})\f_{3}(x_{3})\f_{4}(x_{4})\>=\bigg(\frac{x_{24}^{2}}{x_{14}^{2}}\bigg)^{\frac{1}{2}\Delta_{12}}\bigg(\frac{x_{14}^{2}}{x_{13}^{2}}\bigg)^{\frac{1}{2}\Delta_{34}}\frac{G(u,v)}{(x_{12}^{2})^{\frac{1}{2}(\Delta_{1}+\Delta_{2})}(x_{34}^{2})^{\frac{1}{2}(\Delta_{3}+\Delta_{4})}}.
\eea
Here $G(u,v)$ is an arbitrary function of the conformal cross ratios
\bea
u=\frac{x_{12}^{2}x_{34}^{2}}{x_{13}^{2}x_{24}^{2}}, \ \ v=\frac{x_{14}^{2}x_{23}^{2}}{x_{13}^{2}x_{24}^{2}}.
\eea
\noindent Next we note that the OPE of two scalars takes the general form
\bea
\f_{1}(x_{1})\f_{2}(x_{2})=\sum_{\mathcal{O}}\lambda_{12\mathcal{O}}C(x_{12},\partial_{x_{2}})^{e_{1}...e_{\ell}}\mathcal{O}_{e_{1}...e_{\ell}}(x_{2}).
\eea
The sum runs over all primary operators which can appear in this OPE. 
The contribution of all the descendants is given by the kinematical function $C(x_{12},\partial_{x_{2}})$, which can be found using the 3-point functions by multiplying both sides with $\mathcal{O}_{f_1\dots f_\ell}(x_3)$ and taking the vacuum expectation value. 
The OPE coefficients, $\lambda_{12\mathcal{O}}$, are related to the coefficient of the 3-point functions and are not determined by kinematics. Applying the OPE for $\f_{1}\f_{2}$ and $\f_{3}\f_{4}$ the contribution of a single irreducible representation is given by a conformal block $g_{\mathcal{O}}(u,v)$ or equivalently the conformal partial wave $W_{\mathcal{O}}(x_{1},x_{2},x_{3},x_{4})$ \cite{DO1,DO2,DO3},
\bea
\label{eqn:2}
\<\f_{1}(x_{1})\f_{2}(x_{2})\f_{3}(x_{3})\f_{4}(x_{4})\> &=& \sum_{\mathcal{O}}\lambda_{12\mathcal{O}}\lambda_{34\mathcal{O}}W_{\mathcal{O}}(x_{1},x_{2},x_{3},x_{4}),
\\
W_{\mathcal{O}}(x_{1},x_{2},x_{3},x_{4}) &=& C(x_{12},\partial_{x_{2}})^{e_{1}...e_{\ell}}C(x_{34},\partial_{x_{4}})^{f_{1}...f_{\ell}}\<\mathcal{O}_{e_{1}...e_{\ell}}(x_{2})\mathcal{O}_{f_{1}...f_{\ell}}(x_{4})\>\nn,
\\
&=&\bigg(\frac{x_{24}^{2}}{x_{14}^{2}}\bigg)^{\frac{1}{2}\Delta_{12}}\bigg(\frac{x_{14}^{2}}{x_{13}^{2}}\bigg)^{\frac{1}{2}\Delta_{34}}\frac{g_{\mathcal{O}}(u,v)}{(x_{12}^{2})^{\frac{1}{2}(\Delta_{1}+\Delta_{2})}(x_{34}^{2})^{\frac{1}{2}(\Delta_{3}+\Delta_{4})}},\nn
\eea
where $g_\mathcal{O}$ and $W_\mathcal{O}$ also depend on the external scaling dimensions\footnote{In this paper we will work with a slightly different normalization than the one found using the above method, namely our conformal blocks will have an extra factor of $(-2)^{\ell}$: $g^{(here)}_{\mathcal{O}}=(-2)^{\ell}g^{OPE}_{\mathcal{O}}$. An extra factor of $(-1/2)^{\ell}$ will then appear multiplying $\lambda_{12\mathcal{O}}\lambda_{34\mathcal{O}}$ in the conformal block expansion.}.

To repeat the same idea for operators with spin we need to consider the OPE
\bea
\phi_{1}^{\{a\}}(x_{1})\phi_{2}^{\{b\}}(x_{2})=\sum_{\cO} \lambda_{12\mathcal{O}}C(x_{12},\partial_{x_{2}})^{\{a,b,e\}}\mathcal{O}_{\{e\}},
\eea
where we have used the shorthand $\{a\}\equiv a_{1}...a_{\ell}$. As described in~\cite{Costa:2011dw}, the OPE structures for spinning operators can be found by acting with differential operators on the scalar structures
\bea
C(x_{12},\partial_{x_{2}})^{\{a,b,e\}}=D^{a,b}_{x_{1},x_{2}}C(x_{12},\partial_{x_{2}})^{\{e\}}.
\eea
$D^{a,b}$ is a differential operator constructed via the methods of Section \ref{subsec:3-pointFunctions}. The conformal partial waves are then given by
\bea\label{eq:SpinningPartialWave}
W_{\mathcal{O}}^{\{a,b,c,d\}}(x_{1},x_{2},x_{3},x_{4})=D^{a,b}_{x_{1},x_{2}}D^{c,d}_{x_{3},x_{4}}W_{\mathcal{O}}(x_{1},x_{2},x_{3},x_{4}).
\eea
To be more explicit, the 4-point function of spinning operators is:
\bea
\label{eqn:1}
\<\Phi(P_{1},Z_{1})\Phi(P_{2},Z_{2})\Phi(P_{3},Z_{3})\Phi(P_{4},Z_{4})\>=X_{s}(\{P_i\})\sum_{k}G^{s}_{k}(u,v)Q^{(k)}_{\chi_{1}\chi_{2}\chi_{3}\chi_{4}}(\{P_{i};Z_{i}\}), 
\\
X_{s}(\{P_i\})=\frac{\big(\frac{P_{24}}{P_{14}}\big)^{\frac{\sigma_{1}-\sigma_{2}}{2}}\big(\frac{P_{14}}{P_{13}}\big)^{\frac{\s_{3}-\s_{4}}{2}}}{(P_{12})^{\frac{\s_{1}+\s_{2}}{2}}(P_{34})^{\frac{\s_{3}+\s_{4}}{2}}}, \
\eea
\noindent where $\chi_{i}$ denotes the representation of the operator and $Q^{(k)}$ denote independent tensor structures. $X_{s}$ is the universal prefactor appearing in the s-channel expansion. The coefficient functions, $G^s_{k}(u,v)$, depend only on the conformal cross ratios. The conformal block decomposition is:  
\be
G^{s}_{k}(u,v)=\sum_{\mathcal{O},i,j}\bigg(\frac{-1}{2}\bigg)^{\ell}\lambda^{(i)}_{\Phi_{1}\Phi_{2}\mathcal{O}}\lambda^{(j)}_{\Phi_{3}\Phi_{4}\mathcal{O}}g^{12,34,(ij)}_{\mathcal{O},k}(u,v).
\ee
Note that an exchanged operator can generically give rise to different tensor structures. Rewriting (\ref{eq:SpinningPartialWave}) in terms of conformal blocks, we get: 
\bea
g^{(12,34),(ij)}_{\mathcal{O}}(u,v)=X_{s}^{-1}\mathcal{D}_{L}^{s,(i)}\mathcal{D}_{R}^{s,(j)}W_{\mathcal{O}}^{\{\Delta_{1},\Delta_{2},\Delta_{3},\Delta_{4}\}}(P_{1},P_{2},P_{3},P_{4}),
\\
g^{12,34,(ij)}_{\mathcal{O},k}(u,v)=g^{12,34,(ij)}_{\mathcal{O}}(u,v)\big|_{k},
\eea 
where $\mathcal{D}_{L}^{s,(i)}$ and $\mathcal{D}_{R}^{s,(j)}$ give the s-channel differential operators and $\big|_{k}$ means we project onto the $k$-th tensor structure. We will construct the differential operators case by case explicitly. We have two extra indices, $i$ and $j$, which label the independent 3-point function tensor structures for each operator. That is, the 3-point function $\<\Phi_{1}\Phi_{2}\mathcal{O}\>$ may have multiple, linearly independent tensor structures with unfixed relative coefficients. For example, in three dimensions, $\<TTT\>$ has three structures, two parity even and one parity odd. The parity even structures can be identified with the structures found in a theory of free bosons or free fermions while the odd structure can only appear in an interacting theory. The superscript labels the OPE channels under consideration. To simplify notation later we will write expressions in terms of the conformal block coefficients
\bea\label{eqn:DefineP}
P^{12,34;(ij)}_\mathcal{O}\equiv \bigg(\frac{-1}{2}\bigg)^{\ell}\lambda^{(i)}_{\Phi_{1}\Phi_{2}\mathcal{O}}\lambda^{(j)}_{\Phi_{3}\Phi_{4}\mathcal{O}}=\bigg(\frac{1}{2}\bigg)^{\ell}\lambda^{(i)}_{\Phi_{1}\Phi_{2}\mathcal{O}}\lambda^{(j)}_{\Phi_{4}\Phi_{3}\mathcal{O}}.
\eea
In all cases under investigation this matrix is diagonal in the differential basis (at leading order in $1/\ell$ where we will be working), and positive definite.

Everything we have said corresponds to doing a conformal partial wave expansion in the (12)-(34) channel or s-channel. The partial wave expansion in the (14)-(32) channel, or t-channel, can be found by exchanging $2\leftrightarrow 4$ in all of these expressions. We will denote these t-channel differential operators with a superscript ``t" to distinguish them from the s-channel differential operators.

\section{Lightcone Bootstrap}
\label{sec:bootstrap}

We will now review the lightcone bootstrap when looking at the correlation function of four scalars (see \cite{Fitzpatrick:2012yx,Komargodski:2012ek} for a more thorough analysis). The important result is that for any CFT in $d>2$ spacetime dimensions, the large spin spectrum contains multi-twist states with a Fock space structure whose anomalous dimension asymptotics are determined by the minimal twist sector of the theory. The twist of an operator is defined as the difference between its conformal dimension and spin, $\tau=\Delta-\ell$. Analogous results hold when we consider correlation functions involving the stress-energy tensor and conserved currents, except the contributions of $T_{\mu\nu}$ and $J_{\mu}$ to the anomalous dimension of these double-twist states can have either sign. Requiring that it be non-positive for the stress-energy tensor yields the $d=3$ conformal collider bounds. We will also see interesting behavior when the contribution of $J$ changes sign, but we are not aware of any pre-existing bounds on the relevant OPE coefficient.

\subsection{Review: Scalar 4-point Functions}

We start by reviewing the basic results of~\cite{Fitzpatrick:2012yx,Komargodski:2012ek} and establishing some notation. Given a 4-point function of scalars, $\langle\phi_{1}(x_1)\phi_{1}(x_2)\phi_{2}(x_3)\phi_{2}(x_4)\rangle$, we can perform the OPE inside the correlation function in three different ways, corresponding to three distinct channels. Requiring that the resulting sums of conformal blocks agree yields the bootstrap equations. For our purpose, we only need the bootstrap equations from the (12)-(34) and the (14)-(32) channels,
\begin{equation}\label{eq:Crossingforf1f2}
\sum_{\cO \in \phi_{1,2} \times \phi_{1,2}} P^{11,22}_{\cO}  g^{11,22}_{\tau,\ell}(u,v) = u^{\Delta_2}v^{-\frac{1}{2}(\Delta_1+\Delta_2)} \sum_{\cO \in \phi_1 \times \phi_2} P^{12,21}_{\cO} g^{12,21}_{\tau,\ell}(v,u),
\end{equation}
where the coefficients $P_{\mathcal{O}}^{ij,kl}$ are related to the OPE coefficients as in (\ref{eqn:DefineP}), we label the conformal blocks $g^{ij,kl}_{\tau,\ell}(u,v)$ by the twist $\tau$ and spin $\ell$ of the exchanged operator, and we work in a normalization such that $g_{\tau,\ell}(u,v) \rightarrow u^{\tau/2} (1-v)^{\ell}$ when $u \rightarrow 0$ and then $v \rightarrow 1$. 
 
In the eikonal (or lightcone) limit of $u \ll v \ll 1$, the conformal blocks in (12)-(34) channel are proportional to $u^{\tau/2}$. Therefore the LHS of (\ref{eq:Crossingforf1f2}) is dominated by the low twist operators: the identity with $\tau=0$, conserved currents with $\tau=d-2$ and scalars with low dimensions. In spacetime dimensions $d>2$, the leading $u$ contribution comes exclusively from the identity operator, yielding the following crossing equation: 
\bea\label{eq:Crossingforf1f2Leadingu}
u^{-\Delta_{2}}v^{\frac{1}{2}(\Delta_{1}+\Delta_{2})}=\sum_{\tau,\ell}P^{12,21}_{\tau,\ell}g^{12,21}_{\tau,\ell}(v,u).
\eea

One puzzle is that the left hand side has the power law singularity $u^{-\Delta_2}$ while the crossed channel partial waves can have at most a $u^{\Delta_{1}-\Delta_{2}}$ divergence, for generic $\Delta_{i}$. This problem is even more dramatic if $\Delta_{1}=\Delta_{2}$, in which case the right hand side has at most a $\log(u)$ divergence. The resolution discovered in~\cite{Fitzpatrick:2012yx,Komargodski:2012ek} is that the correct power law singularity can only be reproduced on the RHS by the infinite sum over large spin operators with twist $\Delta_1+\Delta_2$ with the OPE coefficients given by the generalized free field theory. Solving (\ref{eq:Crossingforf1f2Leadingu}) at leading order in $u$ and all orders in $v$ reveals the existence of large spin operators with twists $\Delta_1+\Delta_2+2n$, where $n$ is a non-negative integer. We refer to them as double-twist operators. Solving (\ref{eq:Crossingforf1f2}) to the next leading order in $u$ includes contributions to the LHS from conserved currents and low dimensional scalars, which are reproduced on the RHS by large-$\ell$ suppressed anomalous dimensions of the double-twist operators correcting the canonical twists given above.

To see this explicitly we need an approximate form of the conformal blocks in this limit. Generalizing for the moment, we will start with the conformal block in the (14)-(32) channel when all four operators are distinct scalars with dimensions $\Delta_{i}$. Then in the limit $u\ll v<1$ with $\sqrt{u}\ell\lesssim \mathcal{O}(1)$ we can use the approximation
\bea\label{eq:LargeLConformalBlock}
g_{\tau,\ell}^{12,34}(v,u)\approx\frac{2^{\tau+2\ell}\sqrt{\ell}}{\sqrt{\pi}}u^{\frac{1}{4}(\Delta_{1}+\Delta_{2}-\Delta_{3}-\Delta_{4})}v^{\frac{\tau}{2}}K_{\frac{1}{2}(\Delta_{1}+\Delta_{2}-\Delta_{3}-\Delta_{4})}(2\ell\sqrt{u}),
\eea
where $K_{n}(x)$ is the modified Bessel function of the second kind. Details about the derivation of this equation can be found in Appendix \ref{sec:Appendix_A}. 

The $n=0$ operators, i.e. those with twist $\Delta_{1}+\Delta_{2}$, are required to match the $v^{\frac{1}{2}(\Delta_{1}+\Delta_{2})}$ term on the left hand side. The generalized free field theory OPE coefficients squared in the large spin limit are given by
\bea
\label{eqn:3}
P_{\tau_{0},\ell}\approx \frac{4\sqrt{\pi}}{\Gamma(\Delta_{1})\Gamma(\Delta_{2})2^{\tau_{0}+2\ell}}\ell^{\Delta_{1}+\Delta_{2}-\frac{3}{2}}.
\eea
\noindent After setting $\Delta_{2}=\Delta_{1}$ and $\Delta_{4}=\Delta_{3}$ in the above formula for the conformal block and approximating the sum over $\ell$ as an integral we obtain
\bea
\sum_{\tau,\ell}P^{12,21}_{\tau,\ell}g^{12,21}_{\tau,\ell}(v,u)\approx\frac{4}{\Gamma(\Delta_{1})\Gamma(\Delta_{2})}\int \mathrm{d}\ell \ell^{(\Delta_{1}+\Delta_{2}-1)}K_{\Delta_{1}-\Delta_{2}}(2\ell\sqrt{u}).
\eea
Using the integral
\bea \label{eqn:int}
\int \mathrm{d}\ell\ell^{a}K_{b}(\ell)=2^{a-1}\Gamma\bigg(\frac{1}{2}(1+a-b)\bigg)\Gamma\bigg(\frac{1}{2}(1+a+b)\bigg),
\eea
we reproduce exactly $u^{-\Delta_{2}}v^{\frac{1}{2}(\Delta_{1}+\Delta_{2})}$. Considering a general CFT in $d>2$ dimensions, where we have isolated the identity by taking the $u\rightarrow0$ limit, this illustrates that at large spin there exist double-twist states of the schematic form $\f_{1}\partial_{\mu_{1}}...\partial_{\mu_{\ell}}\f_{2}$\footnote{This form is schematic since we also have to symmetrize the Lorentz indices, remove the traces and the descendant contributions to construct a conformal primary operator. The exact form of these primaries will not be important to us.} whose twists are approximately $\Delta_{1}+\Delta_{2}$. However, if we only have this tower of operators, the crossing equations cannot be solved because their conformal blocks gives higher order contributions in $v$ that do not appear on the left hand side. To cancel these we must include operators with twist $\tau=\Delta_{1}+\Delta_{2}+2n$. These correspond to the  $\f_{1}(\partial^{2})^{n}\partial_{\mu_{1}}...\partial_{\mu_{\ell}}\f_{2}$ operators.

Going to higher order in $u$ in (\ref{eq:Crossingforf1f2}), we must include non-identity operators $\mathcal{O}_m$ with small twist $\tau_{m}$ in the s-channel:
\bea
1+\sum_{\ell_{m}}^{2}P^{11,22}_{m}g^{11,22}_{\tau_{m},\ell_{m}}(u,v)\approx u^{\Delta_2}v^{-\frac{1}{2}(\Delta_1+\Delta_2)} \sum_{\tau,\ell}P_{\tau,\ell}^{12,21}g^{12,21}_{\tau,\ell}(v,u).
\eea
Using this equation, the anomalous dimensions and correction to the OPE coefficients of the double-twist states can be solved in terms of $\tau_m, \ell_m$ and $P^{11,22}_m$. For the moment we will assume that for each $\tau_{n}=\Delta_{1}+\Delta_{2}+2n$ there exists a unique operators at each spin $\ell$, with twist approaching $\tau_{n}$ as $\ell \rightarrow\infty$. We will relax this assumption later when considering 4-point functions of conserved currents.

In a unitary theory the stress energy tensor will always be present in the s-channel with $\tau=d-2$. There is also the possibility of conserved currents with $\tau=d-2$ and scalars with $\frac{d-2}{2}<\tau\leq d-2$. Higher spin conserved currents also have $\tau=d-2$, but the existence of a single higher spin current in a theory with a finite central charge $C_{T}$ would imply the theory is free \cite{Maldacena:2011jn}. Therefore we will restrict the sum on the LHS to $\ell_{m}\leq2$. 

When $u\ll1$ the conformal blocks in the s-channel have the following behavior \cite{DO3}
\bea
g^{12,34}_{\tau,\ell}(u,v)\approx u^{\frac{\tau}{2}}(1-v)^{\ell}\ _{2}F_{1}\bigg(\frac{1}{2}(\tau+\Delta_{2}-\Delta_{1})+\ell,\frac{1}{2}(\tau+\Delta_{3}-\Delta_{4})+\ell,\tau+2\ell;1-v\bigg).
\eea
When $\Delta_{3}=\Delta_{4}$ and $\Delta_{1}=\Delta_{2}$,
\bea
_{2}F_{1}(\beta,\beta,2\beta,1-v)=\frac{\Gamma(2\beta)}{\Gamma^{2}(\beta)}\sum_{n=0}^{\infty}\bigg(\frac{(\beta)_{n}}{n!}\bigg)^{2}v^{n}\bigg(2(\psi(n+1)-\psi(\beta))-\log(v)\bigg),
\eea
where $(x)_{n}=\frac{\Gamma(x+n)}{\Gamma(x)}$ and $\psi(x)=\frac{\Gamma'(x)}{\Gamma(x)}$. 
At leading order in $v$, the $\log(v)$ singularity from the $_{2}F_{1}$ is reproduced by the anomalous dimensions of the double-twist operators on the right hand side: $\tau\rightarrow\Delta_{1}+\Delta_{2}+\gamma(n,\ell)$, where $\gamma(n,\ell)$ is power-law suppressed at large $\ell$. The $\log(v)$ piece arises from expanding $g^{12,21}_{\tau,\ell}(v,u) \sim v^{\frac{\tau}{2}}$ to leading order in the anomalous dimension, $\frac{\gamma(n,\ell)}{2}v^{\frac{\tau}{2}}\log(v)$. The power law singularity in $u$ is then reproduced from the t-channel expansion via the infinite sum over spins. Approximating the anomalous dimension at large $\ell$ as $\gamma(n,\ell)=\frac{\gamma_{n}}{\ell^\delta}$ and matching the $u$ divergence from the s-channel yields $\delta=\tau_{m}$. Matching the $v^{0} \log(v)$ term in the s-channel yields the coefficient \cite{Fitzpatrick:2012yx,Komargodski:2012ek,Fitzpatrick:2014vua}
\bea
\gamma_{0}=-\frac{2\Gamma(\Delta_{1})\Gamma(\Delta_{2})}{\Gamma(\Delta_{1}-\frac{\tau_{m}}{2})\Gamma(\Delta_{2}-\frac{\tau_{m}}{2})}\sum_{\ell_{m}}P_{m}\frac{\Gamma(\tau_{m}+2\ell_{m})}{\Gamma^{2}(\frac{\tau_{m}}{2}+\ell_{m})}.
\eea
The case of matching $\gamma_{0}$ is particularly simple since we just need to multiply each 0th order OPE coefficient by $\frac{1}{2}\frac{\gamma_{0}}{\ell^{\tau_{m}}}$. The t-channel OPE coefficient receives a correction of the form $P_{\tau,\ell} \times c_{n}\ell^{-\tau_{m}}$, where the first coefficient $c_0$ is proportional to $\gamma_0$ and can be found by matching the $v^{0}$ term. Recently there has been success in finding $\gamma_{n}$ for general $n$ in arbitrary dimension \cite{Kaviraj:2015cxa, Kaviraj:2015xsa}, but here we will restrict ourselves to the $\gamma_{0}$ terms.

\subsection{Spinning Operators}\label{subsec:BootstrapSpinningOperators}

The case of external operators with spin is complicated by the presence of multiple tensor structures appearing in the 4-point function. 
Each independent tensor structure yields an independent crossing equation. For the 4-point function of two pairs of identical operators $\<\Phi_{1}(P_{1},Z_{1})\Phi_{1}(P_{2},Z_{2})\Phi_{2}(P_{3},Z_{3})\Phi_{2}(P_{4},Z_{4})\>$, crossing symmetry becomes 
\begin{align}
\label{eqn:spinning}
u^{-\sigma_{2}}v&^{\frac{\sigma_{1}+\sigma_{2}}{2}}G^{s}_{k}(u,v)=G^{t}_{k}(v,u) \ \ \ \forall k,  \\  \nonumber\\
G^{s}_{k}(u,v) &=\sum_{\mathcal{O},i,j} P^{11,22;(ij)}_\mathcal{O} g^{11,22,(ij)}_{\mathcal{O},k}(u,v),
\\
G^{t}_{k}(v,u) &=\sum_{\mathcal{O},i,j}P^{12,21;(ij)}_\mathcal{O}g^{12,21,(ij)}_{\mathcal{O},k}(v,u),
\end{align}
where $k$ runs over the allowed 4-point tensor structures and $\sigma_i = \Delta_i + \ell_i$.

The strategy for solving these equations in the lightcone limit mimics the scalar case. We consider the limit $u\ll v<1$ and $\ell\sqrt{u}\lesssim \mathcal{O}(1)$. The s-channel is dominated in this limit by the operators with minimal twist, which is the identity for $d>2$. The identity contribution has a power law divergence of $u^{-\s_{2}}$, while all the conformal blocks in the t-channel have a weaker power law singularity in $u$. We will show that the identity contribution in the s-channel is reproduced in the t-channel via an infinite sum over spins for multiple families of double-twist states. 
The simplest such states have the schematic form $\Phi_{1,(\rho_{1}...\rho_{\ell_{1}}}\partial_{\mu_{1}}...\partial_{\mu_{k}}\Phi_{2,\nu_{1}...\nu_{\ell_{2}})}$ and have twist $\tau=\tau_{1}+\tau_{2}$ and spin $\ell_{1}+\ell_{2}+k$. But there are other families of double-twist operators arising from contractions between the fields, derivatives, and/or the epsilon tensor. In particular, to reproduce the full tensor structure of the identity exchange in the s-channel, we need to include a few double-twist families with non-minimal twist in the t-channel. The matching will fix their OPE coefficients at leading order in $1/\ell$. 

Unlike in the scalar case we do not know the closed form expressions for the generalized free field theory OPE coefficients. Instead we will make the ansatz that the modified OPE coefficients squared for the double-twist states, the $P^{(ij)}$ terms, have the form $A\ell^{B}2^{-2\ell}$ at large $\ell$. Here A and B are determined by matching the t-channel expansion to the identity contribution. There are a few justifications for this form, besides the fact that it gives the right answer. One is that we computed the exact generalized free theory OPE coefficients for the $[J\phi]_{n,\ell}$ double-twist states using conglomeration \cite{Fitzpatrick:2011dm}, and their large spin limit is of this form. Alternatively, if we follow the analysis of \cite{Fitzpatrick:2015qma}, we can decompose operators with spin into representations of the lightcone (collinear) subgroup of the conformal group. The problem then reduces to a 2d CFT problem where a single correlation function containing an operator with spin can be rewritten in terms of multiple correlation functions containing the lightcone primaries. Then the lightcone bootstrap equations can be solved in the usual way with the scalar collinear blocks and the OPE coefficients take the above form.

With the results from the identity matching, we can expand the crossing equations to the next leading order in $u$ and solve for the large $\ell$ asymptotics of the double-twist anomalous dimensions. In the s-channel, this includes the next-to-minimal twist operators, which we will assume to be conserved spin 1 and 2 operators, whose conformal blocks have an additional $\log(v)$ divergence. 
This logarithm is reproduced in the t-channel in the same way as the scalar case via the anomalous dimensions: $v^{\frac{\tau}{2}}\rightarrow v^{\frac{1}{2}(\tau+\gamma(n,\ell))}\approx \frac{1}{2}\gamma_{n,\ell}v^{\frac{\tau}{2}}\log(v)$. The anomalous dimension asymptotics take the form
\bea
\gamma(n,\ell)=\frac{\gamma_{n}}{\ell^{\delta}},
\eea
where $\delta$ is fixed by reproducing the $u$ dependence in the s-channel. We will find $\delta=1$, which in 3 dimensions is the twist of conserved currents. The coefficient $\gamma_{n}$ can be determined in terms of the s-channel OPE coefficients using the crossing equations.

In our analysis we will only consider contributions in the s-channel from 3-point structures of even parity. In a 3d CFT with a parity symmetry, conserved currents and the stress tensor have even parity so these are guaranteed to be the only contributions. The effect of parity odd contributions, which would arise in theories without parity such as Chern-Simons-matter theories, as well as the effects of scalar exchange, will be considered in future work. 

\subsection{Bootstrap with $\SU(N)$ Adjoint Operators}\label{subsec:SUNBootstrap}

In this subsection, we briefly review the structure of the bootstrap equations when all four operators transform under a global symmetry. In general, the 4-point function will contain several different index structures. Matching their coefficients in different OPE channels generically leads to independent crossing equations. 

We take the adjoint representation of $\SU(N)$ as an example. We will first discuss the scalar 4-point function  $\<\f^a\f^{b}\f^{c}\f^{d}\>$ before generalizing to the spinning case. See \cite{Berkooz:2014yda,Li:2015rfa} for more thorough reviews of the conformal bootstrap with adjoint scalars.

For $N\geq4$, there are 7 representations that can appear in the product of two adjoints (using the notation of \cite{Berkooz:2014yda}): $\big(I\,, Adj_a\,, Adj_s\,, (S,\bar{A})_a \oplus (A,\bar{S})_a\,, (A,\bar{A})_s\,, (S,\bar{S})_s\big)$. 
The subscript $s$ or $a$ denotes whether the operator appears in the symmetric or antisymmetric combination of the two adjoints. The notation $(A,\bar{S})$ means the tensor is antisymmetric in the two fundamental indices and symmetric in the anti-fundamental indices. $(A,\Bar{S})_a$ and $(S,\bar{A})_a$ are complex conjugate and appear together in 4-point functions of real operators. 

The 4-point function can then be decomposed into $6$ independent index structures corresponding to these representations. If we construct them using the   tensor product in the s-channel, we obtain: 
\be
\<\f^a(x_1)\f^b(x_2)\f^c(x_3)\f^d(x_4)\> = X_s \displaystyle{\sum_\bold{r}} G^{s}_\bold{r}(u,v) (I^{s}_\bold{r})^{abcd},
\ee
where $\bold{r} = (I\,, Adj_a\,, Adj_s\,, (S,\bar{A})_a \oplus (A,\bar{S})_a\,, (A,\bar{A})_s\,, (S,\bar{S})_s)$ runs over the $6$ representations and $I^{s}_\bold{r}$ gives the corresponding index structures. We can further decompose each $G^{s}_\bold{r}(u,v)$ in terms of conformal blocks, 
\begin{align}\label{eqn:adjointexpansion}
&G^{s}_{\bold{r}}(u,v) = \sum_{\mathcal{O}\in(\f\times\f)_{\bold{r}}}P_{\mathcal{O}}g_{\Delta_{\mathcal{O}},\ell_{\mathcal{O}}}(u,v), \ \ P_{\mathcal{O}}\equiv\frac{|\lambda_{\mathcal{O}}|^{2}}{2^{\ell_{\mathcal{O}}}}.
\end{align}
We can do a similar decomposition in the t-channel and require that the results agree. Matching the coefficients of the $6$ index structures gives rise to $6$ crossing equations:
\bea
\left(\frac{u}{v}\right)^{\Delta_{\f}} G^{t}_{\bold{r}}(v,u) = \mathcal{M}_\bold{r}^{\bold{r}'} G^{s}_{\bold{r}'}(u,v).
\eea
\noindent The explicit matrix $\mathcal{M}_\bold{r}^{\bold{r}'}$ is given in Appendix \ref{sec:Appendix_C}.

We can generalize this discussion to 4-point functions of operators with spin. Each different choice of global index structure and tensor structure, $(\bold{r},k)$, gives rise to a crossing equation:
\begin{align}
u^{\s_{2}}&v^{-\frac{1}{2}(\s_{1}+\s_{2})}G^{t}_{\bold{r},k}(v,u) =  \mathcal{M}_\bold{r}^{\bold{r}'} G^{s}_{\bold{r}',k}(u,v), \\ \nonumber\\
G^{s}_{\bold{r},k}(u,v) &= \sum_{\mathcal{O}\in(\f\times\f)_{\bold{r}},i,j}(-1)^{\ell_{\bold{r}}}P^{12,34; (ij)}_{\mathcal{O}}g^{(ij)}_{\Delta_{\mathcal{O}},\ell_{\mathcal{O}},k}(u,v), \label{eqn:SU(N)_rules}
\\
G^{t}_{\bold{r},k}(u,v) &= \sum_{\mathcal{O}\in(\f\times\f)_{\bold{r}},i,j}(-1)^{\ell_{\bold{r}}}P^{14,32; (ij)}_{\mathcal{O}}g^{(ij)}_{\Delta_{\mathcal{O}},\ell_{\mathcal{O}},k}(u,v), 
\end{align}
where $\ell_{\bold{r}}$ keeps track of the extra minus signs due to the exchange symmetries of the global index structure. It is 0 for representations that appear in the symmetric product of adjoints and 1 for those that appear in the antisymmetric product of adjoints. Note that here the coefficients $P^{12,34; (ij)}$ also contain a factor of $(-1)^{\ell}$ which in some cases cancels against the $(-1)^{\ell_{\bold{r}}}$ (this implicitly occurred in (\ref{eqn:adjointexpansion})).

For $N=2$ and $3$ some representations do not appear, but the large $\ell$ OPE coefficients and anomalous dimensions can still be found by dropping these representations and setting $N$ to the appropriate values. For $N=3$ the $(A,\bar{A})_{s}$ representation does not appear and for $N=2$ the $(S,\bar{A})_{a}$, $(Adj)_{s}$, and $(A,\bar{A})_{s}$ representations do not appear.

\section{Mixed Current-Scalar 4-point Functions}
\label{sec:JJOO}

In this section we will investigate correlation functions of the form $\<J_{\mu}J_{\nu}\phi\phi\>$, where $J_{\mu}$ is a conserved spin one current and $\phi$ is a real scalar operator of arbitrary dimension. The current has dimension $\Delta_{J}=d-1$ and corresponds to either a $\U(1)$ or $\SU$(N) global symmetry. For the $\U(1)$ case we will take the scalars to be uncharged under the $\U(1)$ symmetry, while for the $\SU$(N) case we will assume they transform in the adjoint representation. There is no loss in generality for the U(1) case since the 3-point function of identical $\U(1)$ currents, $\<JJJ\>$, vanishes for a 3d CFT~\cite{Giombi:2011rz}. 

\subsection{Identity Matching}
\subsubsection{U(1)}

At leading order in the small $u$ limit, the 4-point function is given by: 
\bea
\<J(P_{1};Z_{1})J(P_{2};Z_{2})\phi(P_{3})\phi(P_{4})\>=C_{J}\frac{H_{12}}{(P_{12})^{d}(P_{34})^{\Delta_{\f}}} + \ldots.
\eea
This is the result of the identity exchange in the s-channel. In other words, the 4-point function factorizes at this order and is equal to the generalized free field theory result, even when we are not assuming a large $N$ limit.  $C_J$ is the current central charge and describes the normalization of the current 2-point function
\bea
\<J(P_{1};Z_{1})J(P_{2};Z_{2})\>=C_{J}\frac{H_{12}}{(P_{12})^{d}}.
\eea
In this subsection we reproduce this contribution from the t-channel conformal block expansion. We will show that this requires the t-channel to receive contributions from two families of double-twist operators, the parity even ones $[J\f]_{n,\ell}\sim J_{\nu}(\partial^{2})^{n}\partial_{\mu_{1}}...\partial_{\mu_{\ell-1}}\phi$ with twist $\tau_{n,\ell}=\tau_{J}+\tau_{\f}+2n$, as well as the parity odd ones $[\widetilde{J\f}]_{n,\ell}\sim \epsilon^{\kappa\nu\rho}J_{\nu}\partial_{\rho}(\partial^{2})^{n}\partial_{\mu_{1}}...\partial_{\mu_{\ell-1}}\phi$ with twist $\tau_{J}+\Delta_{\f}+2n+1$. We will solve the crossing equations at leading order in $u$ and $v$, where only the lowest twist states ($n=0$) from both families contribute. 

To construct the spinning conformal blocks in the t-channel, we use the fact that the $\<J\f[J\f]\>$ 3-point function can be represented in terms of differential operators acting on the scalar correlator. After imposing the conservation conditions, for $[J\f]_{0,\ell}$ we have
\be
\<J(P_{1};Z_{1})\f(P_{2})[J\f]_{0,\ell}(P_{3};Z_{3})\>=\big(\frac{2-d}{\ell+\Delta_{\f}-1}D_{11}\Sigma_{L}^{1,0}+D_{12}\Sigma_{L}^{0,1}\big)\frac{\hat{\lambda}_{J\f[J\f]_{0,\ell}}V_{3}^{\ell}}{(P_{12})^{\frac{1}{2}-\ell}(P_{13})^{d+\ell-\frac{3}{2}}(P_{23})^{\Delta_{\f}+\ell-\frac{1}{2}}}, \nonumber\\
\ee
where $\hat{\lambda}_{J\f[J\f]_{0,\ell}}$ is an arbitrary coefficient. For the parity odd double-twist states, we have
\bea
\<J(P_{1};Z_{1})\phi(P_{2})[\widetilde{J\f}]_{0,\ell}(P_{3};Z_{3})\>=\tilde{D}_{1}\frac{\hat{\lambda}_{J\f\widetilde{[J\f]}_{0,\ell}}V_{3}^{\ell}}{(P_{12})^{\frac{1}{2}-\ell}(P_{13})^{d+\ell-\frac{3}{2}}(P_{23})^{\Delta_{\f}+\ell-\frac{1}{2}}}.
\eea
\noindent This correlation function is automatically conserved. 
The t-channel differential operators that generate the spinning blocks are then\footnote{We explicitly label the differential operators by $t$ as a reminder that these are the differential operators for the t-channel. In other words, in our original formulas we must let $2\rightarrow4$. In these expressions $\Sigma_{L}^{a,b}$ shifts $\Delta_{1}$ by $a$ and $\Delta_{4}$ by $b$, while $\Sigma_{R}^{a,b}$ shifts $\Delta_{2}$ by $a$ and $\Delta_{3}$ by $b$.}
\begin{align}
&D_{[J\f]_{0,\ell}}^{t}=\bigg(-\frac{1}{\ell+\Delta_{\f}-1}D^{t}_{11}\Sigma^{1,0}_{L,t}+D^{t}_{14}\Sigma^{0,1}_{L,t}\bigg)\bigg(-\frac{1}{\ell+\Delta_{\f}-1}D^{t}_{22}\Sigma^{1,0}_{R,t}+D^{t}_{23}\Sigma^{0,1}_{R,t} \bigg),\\
&D_{\widetilde{[J\f]}_{0,\ell}}^{t}=\tilde{D}^{t}_{1}\tilde{D}^{t}_{2},
\end{align}
\noindent where we have set $d=3$. 
\noindent The crossing equation at leading order in $u$ is
\bea\label{eqn:JJffCrossing}
C_{J}\frac{H_{12}}{(P_{12})^{3}(P_{34})^{\Delta_{\f}}}=&\sum_{n,\ell}P_{[J\f]_{n,\ell}}D_{[J\f]_{n,\ell}}^{t}W_{[J\f]_{n,\ell}}^{t}+P_{[\widetilde{J\f}]_{n,\ell}}D_{\widetilde{[J\f]}_{n,\ell}}^{t}W_{\widetilde{[J\f]}_{n,\ell}}^{t}, \ \
\eea
where $P_{[J\f]_{n,\ell}}$ and $P_{[\widetilde{J\f}]_{n,\ell}}$ are positive squares of OPE coefficients as normalized in (\ref{eqn:DefineP}). $W_{\mathcal{O}}^{t}$ denotes the t-channel conformal partial wave with scalars, which is (\ref{eqn:2}) with $2\leftrightarrow4$ exchanged. 

We now solve (\ref{eqn:JJffCrossing}) at the lowest order in $v$. Thus we set $n=0$ and restrict the sum to be over $\ell$.\footnote{The crossing equation only depends on the cross ratios $u,v$ even when this is not manifest in (\ref{eqn:JJffCrossing}).} 
As mentioned in Section \ref{subsec:BootstrapSpinningOperators}, we make the ansatz that as $\ell\rightarrow\infty$ we have $P_{i}\approx A_{i}\ell^{B_{i}}2^{-2\ell}$. The bootstrap equation is now straightforward solve, we act with the differential operators on the large spin conformal blocks in (\ref{eq:LargeLConformalBlock}), which produces terms of a similar form, $\sim \ell^{a}K_{b}(2\ell\sqrt{u})$. The sum is then approximated as an integral and evaluated using (\ref{eqn:int}). Computationally, it is more efficient to compute the integral first and then act with the differential operators. The result gives the OPE coefficients at leading order in $1/\ell$:
\bea
P_{[J\f]_{0,\ell}}\approx \frac{\sqrt{\pi}C_{J}}{2^{2\ell+\Delta_{\f} -1}\Gamma(\Delta_{\f})}\ell^{\frac{1}{2}(2\Delta_{\f}-1)}, \hspace{1cm} P_{[\widetilde{J\f}]_{0,\ell}}\approx \frac{\sqrt{\pi}C_{J}}{2^{2\ell+\Delta_{\f}+1}\Gamma(\Delta_{\f})}\ell^{\frac{1}{2}(2\Delta_{\f}-7)}.
\eea

\subsubsection{$\SU(N)$}\label{subsec:JJffIdentitySUN}

At leading order in $u$, the 4-point function is dominated by the identity exchange:
\bea
\<J^a(P_{1};Z_{1})J^b(P_{2};Z_{2})\phi^c(P_{3})\phi^d(P_{4})\>=C_{J}\delta^{ab}\delta^{cd}\frac{H_{12}}{(P_{12})^{d}(P_{34})^{\Delta_{\f}}} + \ldots.
\eea
As explained in Section \ref{subsec:SUNBootstrap}, there are 6 independent bootstrap equations corresponding to the 6 index structures. They are given in Eq.~(\ref{eqn:SU(N)_rules}) with the matrix defined in Appendix \ref{sec:Appendix_C}. For each equation the analysis is the same as the $\U(1)$ case. Matching the identity contribution shows that double-twist operators in all representations appear in the t-channel with the following coefficients:
\begin{align}
\bold{P}_{[J\f]_{0,\ell}}=\frac{C_{J}\sqrt{\pi }}{2^{\Delta_{\f}+1+2\ell}\Gamma(\Delta_{\f})} \ell^{\frac{1}{2}(2\Delta_{\f}-1)} \bold{P}, \hspace{1cm}
\bold{P}_{[\widetilde{J\f}]_{0,\ell}}=\frac{C_{J}\sqrt{\pi}}{2^{\Delta_{\f}+3+2\ell}\Gamma (\Delta_{\f})}\ell^{\frac{1}{2}(2\Delta_{\f}-7)}\bold{P},
\end{align}
\noindent where we have defined the vector $\bold{P}=\big(\frac{4}{N^2-1},\frac{2}{N},\frac{2 N}{N^2-4},2,1,1\big)$ using the basis of representations $\big(I\,, Adj_a\,, Adj_s\,, (S,\bar{A})_a \oplus (A,\bar{S})_a\,, (A,\bar{A})_s\,, (S,\bar{S})_s\big)$.
One interesting point is that the OPE coefficients for the singlet representations decay like $\sim 1/N^{2}$, the $(Adj)_{s}$ and $(Adj)_{a}$ coefficients like $\sim 1/N$, and all others approach a constant as $N\rightarrow\infty$, showing that the former states decouple at large $N$.

\subsection{Stress Tensor and Current Matching}

\subsubsection{U(1)}

We now solve the crossing equation at the next leading order in $u$. In the $\U(1)$ case, the s-channel contains the contribution of the stress-energy tensor.\footnote{To simplify the analysis, we are assuming that there are no scalars with $\frac{1}{2}<\Delta<1$.  Their contributions can also be included with the methods described here.}
The spinning conformal block for stress-tensor exchange can be obtained by acting with a differential operator $D_{L,T}$ on the scalar partial wave, which is fixed by the condition $D_{L,T}\<\f_J \f_J T\>=\<JJT\>$, where in general $\f_J$ is a real scalar with dimension $d-1$. Conformal symmetry implies
\bea
\<J(P_{1};Z_{1})J(P_{2};Z_{2})T(P_{3};Z_{3})\>=\frac{\alpha V_{1}V_{2}V_{3}^{2}+\beta(H_{13}V_{2}+H_{23}V_{1})V_{3}+\gamma H_{12}V_{3}^{2}+\eta H_{13}H_{23}}{(P_{12})^{\frac{d}{2}-1}(P_{13})^{\frac{d}{2}+1}(P_{23})^{\frac{d}{2}+1}} .\nonumber
\\
\eea
 After imposing conservation conditions and the Ward identity, this 3-point function is fixed in terms of two coefficients, $\lambda_{JJT}$ and $C_J$~\cite{Osborn:1993cr}:
 \begin{align}
 \alpha &= d\frac{C_J( d^2 - 4)-2\lambda_{JJT} d S_d}{2 S_d},\hspace{1cm} \beta = -2\lambda_{JJT}, \\
 \gamma &= -d \lambda_{JJT} +\frac{d(d-2) C_J}{2 S_d},\hspace{1.6cm} \eta = -2\lambda_{JJT} + \frac{d C_J}{S_d},
 \end{align}
where $C_J$ is the current central charge and $S_{d}$ is the volume of a $d-1$-dimensional sphere, \begin{math} S_{d}=\frac{2\pi^{\frac{d}{2}}}{\Gamma(\frac{d}{2})}\end{math}. The coefficient $\lambda_{JJT}$ is arbitrary and was denoted as $c$ in~\cite{Osborn:1993cr} and $\hat{c}$ in~\cite{Chowdhury:2012km}.

We can reproduce this 3-point function with the following differential operator: 
\bea\label{eqn:DLT}
D_{L,T}=\bigg[\bigg(2\lambda_{JJT}-\frac{d(d-2)C_{J}}{(d-1)S_{d}}\bigg)D_{11}D_{22}+\bigg(2\lambda_{JJT}-\frac{d^2 C_{J} }{(d-1)S_{d}}\bigg)D_{12}D_{21}-2\lambda_{JJT} H_{12}\bigg]\Sigma_{L}^{1,1}\label{eqn:JJT}, \hspace{0.5cm}
\eea
\noindent where $\Sigma^{a,b}_{L}$ shifts the dimensions of the first two operators. 

In the 4-point function, the identity contribution is corrected by the stress tensor exchange in the s-channel, which is suppressed by an extra factor of $\sqrt{u}$:
\bea
\<J(P_{1};Z_{1})J(P_{2};Z_{2})\phi(P_{3})\phi(P_{4})\>=C_{J}\frac{H_{12}}{(P_{12})^{d}(P_{34})^{\Delta_{\f}}} + \lambda_{\f\f T}\frac{1}{4\sqrt{C_T}}D_{L,T}W^{s}_{T}(\{P_{i}\}) + \ldots, \ \,
\eea
where $\lambda_{\f\f T}$ is also fixed by the Ward identity to be
\be
\lambda_{\f\f  T}=-\frac{\Delta_{\f}d}{(d-1)S_{d}}\frac{1}{\sqrt{C_{T}}}.
\ee
In this first correction, the leading contribution at small $v$ is a logarithmic singularity. This $\log(v)$ is matched in the t-channel by expanding the conformal blocks in the anomalous dimensions of double-twist operators. The crossing equation then takes the form\footnote{The division by $\sqrt{C_{T}}$ comes from how we normalize the stress energy tensor. A division by $\sqrt{C_{J}}$ will also appear for current exchange. See Appendix \ref{App:3pfsDiffOps} for our conventions.}  
\begin{align}
\frac{\lambda_{\f\f T}}{4\sqrt{C_{T}}}D_{L,T}W^{s}_{T}=\sum_{n, \ell}P_{[J\f]_{n,\ell}}\gamma_{[J\f]_{n,\ell}}\partial_{\tau_{n}}D^{t}_{[J\f]_{n,\ell}}W_{[J\f]_{n,\ell}}^{t}+P_{\widetilde{[J\f]}_{n,\ell}}\gamma_{\widetilde{[J\f]}_{n,\ell}}\partial_{\tilde{\tau}_{n}}D^{t}_{\widetilde{[J\f]}_{n,\ell}}W_{\widetilde{[J\f]}_{n,\ell}}^{t},
\end{align}
\noindent where we implicitly restrict to terms multiplying $\log(v)$. This equation can be solved with the same technique as before. The anomalous dimensions are necessarily $1/\ell$ suppressed at large $\ell$ for their effect to appear at the correct order in $u$. With the notation $\gamma_{\mathcal{O}_{n,\ell}} \equiv \gamma_{\mathcal{O}_n}/\ell$, the $n=0$ results are:
\bea\label{eqn:gammaJfU1}
\gamma_{[J\f]_{0}}= -\frac{4 \Delta_{\f} (3 C_{J}-8 \pi  \lambda_{JJT}) \Gamma (\Delta_{\f})}{\pi ^{7/2} C_{T} C_{J} \Gamma \left(\Delta_{\f}-\frac{1}{2}\right)}, \hspace{1cm} 
\gamma_{\widetilde{[J\f]}_{0}}= -\frac{8 \Delta_{\f} (16 \pi  \lambda_{JJT}-3 C_{J}) \Gamma (\Delta_{\f})}{\pi ^{7/2} C_{T} C_{J} \Gamma \left(\Delta_{\f}-\frac{1}{2}\right)}.
\eea

In an AdS bulk description, these anomalous dimensions correspond to the gravitational binding energies for well-separated 2-particle states~\cite{Fitzpatrick:2012yx}. Since gravity is expected to be attractive, we expect the anomalous dimensions to be negative. 
Assuming that the CFT is unitary, so $C_{T}>0$ and $C_{J}>0$, and that the scalar is not free, or equivalently $\Delta_{\f}>\frac{1}{2}$, we see that the anomalous dimensions are non-positive if and only if the relevant 3d conformal collider bounds \cite{Chowdhury:2012km} are satisfied:
\bea\label{eqn:ColliderBoundsForJJT}
\frac{3C_{J}}{16\pi}\leq \lambda_{JJT} \leq\frac{3C_{J}}{8\pi}.
\eea
These bounds were originally discovered by requiring the integrated energy flux at spatial infinity due to a localized perturbation in the CFT to be positive. Our results suggest that the positivity of this energy flux is equivalent to the attractiveness of bulk gravity at long distances. We could also turn this around and conclude that our analysis combined with the conformal collider bounds gives an argument for the attractive nature of bulk gravity at long distances, using entirely properties of the field theory. The negativity of the anomalous dimensions is also related to bulk causality in large $N$ theories~\cite{Camanho:2014apa}. We hope to explore this connection in more detail in future work.

Note that the two boundary values, $\frac{3C_{J}}{8\pi}$ and $\frac{3C_{J}}{16\pi}$, correspond to the values found in a theory of free fermions and free bosons respectively. When $\lambda_{JJT}$ saturates one of the bounds, one of the asymptotic anomalous dimensions vanishes. This could be an indication that certain sectors of the theory are decoupling (see \cite{Zhiboedov:2013opa} for related work in 4d). 
It would be interesting to extend this analysis to higher order in $n$ or $\ell$ to see if this behavior continues to hold. 
The fact that the other anomalous dimensions remains non-zero at the boundary free values indicates that our analysis is incomplete for free theories, which contain an infinite number of higher spin conserved currents. We have not included them in the s-channel analysis since we focused on interacting theories which have a twist gap separating the spin $1,2$ conserved currents from the other operators. After summing up all the contributions from higher spin conserved currents, we expect the logarithm to disappear and the anomalous dimensions of the double-twist states to vanish. 

\subsubsection{$\SU(N)$}\label{subsec:JfJfNLOSUN}
For a 4-point function of $\SU(N)$ adjoint operators, the $\SU(N)$ conserved currents can appear in the OPE decomposition. This gives rise to another contribution at the same order compared to the stress tensor exchange. The spinning conformal block for the current exchange can be obtained by acting with a differential operator $D_{L,J}$ on the scalar partial wave, which is fixed by the condition $D_{L,J}\<\f_J \f_J J\>=\<JJJ\>$, where $\f_J$ is a real scalar with dimension $d-1$. After imposing the conservation conditions and the Ward identity, we see that this 3-point function is fixed by two coefficients, $C_J$ and $\lambda_{JJJ}$~\cite{Osborn:1993cr}: 
\begin{eqnarray}\label{eqn:JJJ}
\<J^{a}(P_{1},Z_{1})J^{b}(P_{2},Z_{2})J^{c}(P_{3},Z_{3})\>=f^{abc}\frac{(\frac{C_J d}{S_d}-(2+d)\lambda_{JJJ})V_{1}V_{2}V_{3} - \lambda_{JJJ}(H_{12}V_{3}+H_{13}V_{2}+H_{23}V_{1})}{(P_{12})^{\frac{d}{2}}(P_{13})^{\frac{d}{2}}(P_{23})^{\frac{d}{2}}}, \nonumber\\ \hspace{-0.3cm}
\end{eqnarray}
\noindent where $f^{abc}$ are the structure constants, corresponding to the $(Adj)_{a}$ representation in the tensor product. In~\cite{Osborn:1993cr} the coefficient $\lambda_{JJJ}$ was called $b$. 

The associated differential operator that generates this 3-point function is given by
\begin{eqnarray}
D_{L,J}=\frac{d C_{J}}{(d-2)S_{d}}(D_{12}D_{22}\Sigma_{L}^{0,2}+D_{11}D_{21}\Sigma_{L}^{2,0}-D_{12}D_{21}\Sigma_{L}^{1,1})+\frac{4\lambda_{JJJ}S_{d}-dC_{J}}{(d-2)S_{d}}D_{11}D_{22}\Sigma_{L}^{1,1}. \hspace{.5cm}  \label{eqn:DJJJ}
\end{eqnarray}
This is all we need to construct the spinning conformal block. After solving the crossing equation at the next-to-leading order in $u$, we find that the t-channel anomalous dimensions receive separate contributions from $T$ and $J$ exchange.
The anomalous dimensions asymptote to $\boldsymbol{\gamma}_{O_{n,\ell}}=\boldsymbol{\gamma}_{O_{n}}\ell^{-1}$ at large $\ell$. The $n=0$ coefficients are given by:
\begin{align}\label{eqn:gammaJphiSUN}
\boldsymbol{\gamma}_{[J\f]_0}=&-\frac{2 (C_{J}-4 \pi  \lambda_{JJJ})\Gamma (\Delta_{\f})}{\pi ^{7/2} C_{J}^2 \Gamma \left(\Delta_{\f}-\frac{1}{2}\right)}\boldsymbol{\gamma}_{J}-\frac{4 \Delta_{\f} (3 C_{J}-8 \pi  \lambda_{JJT}) \Gamma (\Delta_{\f})}{\pi ^{7/2} C_{T} C_{J} \Gamma \left(\Delta_{\f}-\frac{1}{2}\right)}\boldsymbol{\gamma}_{T},\nonumber\\
\boldsymbol{\gamma}_{[\widetilde{J\f}]_0}=&-\frac{2 (8 \pi  \lambda_{JJJ}-C_{J}) \Gamma (\Delta_{\f})}{\pi ^{7/2} C_{J}^2 \Gamma \left(\Delta_{\f}-\frac{1}{2}\right)}\boldsymbol{\gamma}_{J}-\frac{8 \Delta_{\f} (16 \pi  \lambda_{JJT}-3 C_{J}) \Gamma (\Delta_{\f})}{\pi ^{7/2} C_{T} C_{J} \Gamma \left(\Delta_{\f}-\frac{1}{2}\right)}\boldsymbol{\gamma}_{T},\\
\boldsymbol{\gamma}_{J}=&(2N, N, N, 0, 2, -2), \hspace{0.5cm} \boldsymbol{\gamma}_{T}=(1,1,1,1,1,1),\nonumber\label{eqn:gammaJgammaT}
\end{align}
where $\boldsymbol{\gamma}_{J}$ and $\boldsymbol{\gamma}_{T}$ give the results for double-twist operators in different representations of $\SU(N)$: $\big(I\,, Adj_a\,, Adj_s\,, (S,\bar{A})_a \oplus (A,\bar{S})_a\,, (A,\bar{A})_s\,, (S,\bar{S})_s\big)$.  
The second terms in these expressions are the corrections to the dimension due to the stress tensor exchange in the s-channel. 
As in the $\U(1)$ case, they correspond to the gravitational binding energies between well separated 2-particle states in AdS. 
The fact that they are the same for different representations is consistent with the universality of gravity. 
From the CFT perspective this is due to the fact that $T$ appears in the singlet representation of $\SU(N)$.
Since gravity at long distance is expected to be attractive, we expect these anomalous dimensions to be negative. 
Once again, we find that this holds if and only if the same conformal collider bounds (\ref{eqn:ColliderBoundsForJJT}) are satisfied. 

The anomalous dimensions from current exchange are given by the first terms in (\ref{eqn:gammaJphiSUN}). In a dual AdS description, they correspond to the binding energy from non-Abelian gauge interactions for well-separated 2-particle states. We find that this interaction is attractive for the neutral 2-particle states, or the singlet, if and only if
\be\label{eqn:boundsForJJJ}
\frac{C_{J}}{8\pi}\le \lambda_{JJJ} \le \frac{C_{J}}{4\pi},
\ee
where $\lambda_{JJJ}$ and $C_{J}$ parameterize $\<JJJ\>$. In a theory of free bosons $\lambda_{JJJ}=\frac{C_{J}}{8\pi}$ while in a theory of free fermions $\lambda_{JJJ}=\frac{C_{J}}{4\pi}$~\cite{Osborn:1993cr}. Just as in the case of $T$-exchange, when $\<JJJ\>$ saturates the free boson or free fermion structures some of the asymptotic anomalous dimensions vanish. 

The inequalities (\ref{eqn:boundsForJJJ}) are intimately related to conformal collider physics. By acting with a properly chosen non-Abelian current on the vacuum, we can create a localized state with a positive charge under a $\U(1)\subset \SU(N)$ at the origin.\footnote{Although we only analyzed the case of $\SU(N)$, we expect similar features to appear more generally. In particular, from the analysis of \cite{Li:2015rfa} we expect them to show up in 4-point functions of operators in other representations of $\SU(N)$ as well as in CFTs with other global symmetries such as $O(N)$.} This local perturbation propagates and the charge flux at infinity can be measured. We show in  Appendix \ref{App:Charge1PF} that the expectation value of the charge flux at infinity is positive if and only if (\ref{eqn:boundsForJJJ}) is satisfied. 
In contrary to the energy flux, the charge flux in any single event may trivially have different signs at different angles, as expected from a showering of charged particles. But this doesn't imply that (\ref{eqn:boundsForJJJ}) is generically violated. Indeed, to make the expectation value of the charge flux negative, one needs a large charge flux asymmetry. This is the much more non-trivial behavior that is forbidden by (\ref{eqn:boundsForJJJ}). 

From the perspective of gauge interactions in the bulk, the regime violating (\ref{eqn:boundsForJJJ}) seems rather strange. (\ref{eqn:boundsForJJJ}) is equivalent to the statement that the gauge representation of a well separated 2-particle state determines the sign of its gauge binding energy. In particular, this sign will not depend on the spin of the particles or the parity of the state. This follows from comparing (\ref{eqn:gammaJphiSUN}) to the corresponding result in scalar 4-point functions~\cite{Li:2015rfa}, where this sign is uniquely determined by the representation. However if, for example, $\lambda_{JJJ}>\frac{C_{J}}{4\pi}$, then all the parity-even 2-particle states consisting of a scalar and a gauge boson $[J\phi]$ will have binding energies with opposite signs compared to the scalar-scalar state $[\phi\phi]$ or the parity odd states $[\widetilde{J\phi}]$. These behaviors seems counter-intuitive. For example, the singlet 2-particle state $[J\phi]$ which intuitively holds the least energy in the gauge field configurations will become the most energetic one. We do not have a rigorous way to forbid this situation, but it is tempting to conjecture that the bound (\ref{eqn:boundsForJJJ}) holds in all unitary CFTs.

It would be interesting to see if there could exist consistent CFTs or theories in AdS that violate (\ref{eqn:boundsForJJJ}). We are not aware of any examples. Some holographic constraints on massive triple vector boson couplings were found in~\cite{Kulaxizi:2012xp}, but their analysis does not apply to this case. For all superconformal theories this bound holds. Supersymmetry fixes the parity even 3-point functions of conserved global symmetry currents up to an overall coefficient \cite{Nizami:2013tpa}. Therefore, $\lambda_{JJJ}$ can be calculated in a free theory and the result holds for all SCFTs since the positivity of the number of bosons and fermions in the free theory implies (\ref{eqn:boundsForJJJ}). 

Furthermore, using the constraints of slightly broken higher spin symmetry, Maldacena and Zhiboedov \cite{Maldacena:2012sf} found the correlation functions of all currents to leading order in $N$ in two classes of CFTs parametrized by an effective 't Hooft coupling $\lambda$, which they called the quasi-bosonic and quasi-fermionic theories. Large $N$ Chern-Simons theories with fundamental matter fall into this category and include as special cases the critical O(N) model and UV Gross-Neveu O(N) model. Specifically they found that $\<J^{(s_{1})}J^{(s_{2})}J^{(s_{3})}\>$ is parametrized by three structures $\<\>_{bos}$, $\<\>_{fer}$, $\<\>_{odd}$, which refer to correlators found in a theory of free bosons, free fermions, and a structure that only shows up in an interacting theory. The coefficients in front of the bosonic and fermionic structures are always positive semidefinite, so the conjectured bound on $\lambda_{JJJ}$ holds for CFTs with slightly broken higher spin symmetries. 

Finally, let us consider the dependence of the anomalous dimensions with respect to $N$. We expect that $C_{T}$ scales with some positive power of $N$, while the behavior of $C_J$ is less clear.\footnote{The scaling properties of $C_J$ can be determined if we are close to a free field theory description. For example, the contribution of a free field in representation $\bold{r}$ of the global symmetry group to $C_{J}$ scales like the index of the representation $C(\bold{r})$, which is defined as $\Tr_{\bold{r}}(T^{a}T^{b})=C(\bold{r})\delta^{ab}$ \cite{Osborn:1993cr}. For the fundamental representation this is a constant, but for the adjoint representation, this grows like $N$.} The bounds (\ref{eqn:ColliderBoundsForJJT}) and (\ref{eqn:boundsForJJJ}), if true, indicate that $\lambda_{JJT}\sim C_{J}$ and $\lambda_{JJJ}\sim C_{J}$. At large $N$ we see that the contribution of $T$ to the anomalous dimension becomes small for all the operators. This is consistent with bulk gravity turning off. If $C_J$ stays constant as $N\rightarrow\infty$, the anomalous dimensions of double-twist states in the singlet and adjoint channels can start to decrease like $-N/\ell$. In this case our results should only hold when $N\ll \ell$, otherwise we cannot treat the anomalous dimensions as perturbative parameters. If $C_{J}$ also scales with some positive power of $N$, our results may have a wider range of validity. 

\section{Current 4-point Functions}
\label{sec:JJJJ}

In this section we will generalize the above analysis to 4-point functions of currents $\<JJJJ\>$. As before, we will first match the identity contribution in the $s$-channel to an infinite sum over large spin double-twist states. Then we will match the current and stress-tensor contributions to compute the anomalous dimensions of these states.

\subsection{Identity Matching}

\subsubsection{U(1)}

At the leading order in the $u\ll v\ll 1$ limit, the 4-point function factorizes:
\bea\label{eqn:JJJJ4PFLO}
\<J(P_1,Z_1)J(P_2,Z_2)J(P_3,Z_3)J(P_4,Z_4)\>=C_{J}^{2}\frac{H_{12}H_{34}}{(P_{12})^{d}(P_{34})^{d}}+\dots.
\eea
This corresponds to the contribution from the identity exchange in the s-channel.\footnote{There are also identity exchanges in the t- and u-channel, but these contributions are subleading in the small $u$ limit.} We will see that to reproduce all polarizations of (\ref{eqn:JJJJ4PFLO}) at leading order in the lightcone limit from the t-channel conformal block decomposition, we would need to include contributions from the three families of large spin double-twist operators given in Table \ref{table:JJJJDouble-Twists}. It is perhaps surprising that the twist 4 states $[JJ]_{1,\ell}$ should contribute at leading order. One of the polarizations in (\ref{eqn:JJJJ4PFLO}), $(Z_1\cdot Z_2) (Z_3\cdot Z_4)$, receives a contribution at leading order from this state when we take into account degeneracies among the four point function structures (see Appendix \ref{App:DegeneracyEquations}).
\begin{table}
\begin{center}
\begin{tabular}{|c|c|c|c|c|}
\hline 
operator & twist & parity & spin & constructions\tabularnewline
\hline 
\hline 
$[JJ]_{0,\ell}$ & 2 & even & even & $J_{\mu}\partial_{\nu_{1}}...\partial_{\nu_{\ell-2}}J_{\rho}$\tabularnewline
\hline 
$[JJ]_{1,\ell}$ & 4 & even & even & $J_{\mu}\partial^{2}\partial_{\nu_{1}}...\partial_{\nu_{\ell-2}}J_{\rho},\hspace{1em}J^{\mu}\partial_{\nu_{1}}...\partial_{\nu_{\ell}}J_{\mu}$\tabularnewline
\hline 
$[\widetilde{JJ}]_{0,\ell}$ & 3 & odd  & even \& odd & $\epsilon_{\rho}^{\ \mu\kappa}J_{\mu}\partial_{\kappa}\partial_{\nu_{1}}...\partial_{\nu_{\ell-2}}J_{\sigma},\hspace{1em}\epsilon_{\rho}^{\ \mu\kappa}J_{\mu}\partial_{\nu_{1}}...\partial_{\nu_{\ell-1}}J_{\kappa}$\tabularnewline
\hline 
\end{tabular}
\end{center}
\caption{T-channel double-twist operators that reproduce the s-channel identity
exchange in $\langle JJJJ\rangle$. From the bootstrap perspective, we cannot distinguish different constructions with the same twist and parity, with the exception that the second construction of $[\widetilde{JJ}]_{0,\ell}$ exists only for even $\ell$. 
Our solutions represent the sum of their OPE coefficients and the average of their anomalous dimensions, e.g., $P_{\mathcal{O}}\equiv\sum_i P_{\mathcal{O}_i}$ and $\gamma_{\mathcal{O}}\equiv\frac{\sum_i P_{\mathcal{O}_i} \gamma_{\mathcal{O}_i}}{P_{\mathcal{O}_i}}$. }\label{table:JJJJDouble-Twists}
\end{table}

We will start by constructing the spinning conformal blocks associated with these operators. The conformal blocks for exchanging a general spin-$\ell$ operator in $\<JJJJ\>$ can be written in terms of differential operators acting on the known blocks for scalar 4-point functions. In the conformal partial wave expansion, the contribution from a primary $\mathcal{O}_\ell$ to $\<JJJJ\>$ can be written as
\be
P_{\mathcal{O}_\ell}^{(ij)} D_{L,i} D_{R,j} W^{t}_{\mathcal{O}_\ell},
\ee
where $W^{t}_{\mathcal{O}_\ell}$ is the scalar conformal partial wave (\ref{eqn:2}) with $(2\leftrightarrow4)$ permuted. $P^{(ij)}_{\mathcal{O}_\ell}$ are products of OPE coefficients in the normalization of (\ref{eqn:DefineP}). Our goal in this subsection is to solve for them in the large $\ell$ regime with the crossing equations. The t-channel differential operators are~\cite{Costa:2011dw}
\begin{align}
D_{L,1}^{t}=&\bigg(2+\frac{(\Delta-\ell-1)(\Delta-\ell-3)(\Delta+\ell-2)(\Delta+\ell)}{C_{\Delta,\ell}}\bigg)D^{t}_{11}D^{t}_{44}\Sigma_{L}^{1,1}  \nonumber
\\
&-(\Delta-\ell-1)(\Delta+\ell)(D^{t}_{41}D^{t}_{11}\Sigma_{L}^{2,0}+D^{t}_{14}D^{t}_{44}\Sigma_{L}^{0,2})+C_{\Delta,\ell}D^{t}_{14}D^{t}_{41}\Sigma_{L}^{1,1} ,
\\
D^{t}_{L,2}=&-4D^{t}_{11}D^{t}_{44}\Sigma_{L}^{1,1}+C_{\Delta,\ell}H_{14}\Sigma_{L}^{1,1},
\\ \nonumber \\ 
\tilde{D}_{L,+}^{t}=&(D_{41}^{t}\tilde{D}_{1}^{t}\Sigma_{L}^{1,0}+D_{14}^{t}\tilde{D}_{4}\Sigma_{L}^{0,1})+\frac{(3-\Delta) \Delta+\ell(1+\ell)}{C_{\Delta,\ell}}(D_{44}^{t}\tilde{D}_{1}^{t}\Sigma_{L}^{0,1}+D_{11}^{t}\tilde{D}_{4}^{t}\Sigma_{L}^{1,0}),
\\ 
\tilde{D}_{L,-}^{t}=&(D_{41}^{t}\tilde{D}_{1}^{t}\Sigma_{L}^{1,0}-D_{14}^{t}\tilde{D}_{4}^{t}\Sigma_{L}^{0,1})+\frac{(3-\Delta) \Delta+\ell(1+\ell)-4}{C_{\Delta,\ell}}(D_{44}^{t}\tilde{D}_{1}^{t}\Sigma_{L}^{0,1}-D_{11}^{t}\tilde{D}_{4}^{t}\Sigma_{L}^{1,0}),
\end{align}
where $C_{\Delta,\ell}=\Delta(\Delta-d)+\ell(\ell+d-2)$ is the quadratic Casimir and $\Delta,\ell$ refer to the scaling dimension and spin of the exchanged operator. 
Note that each $D^{t}_{L,i}$ respects the conservation conditions of the external currents. The first two are parity even and appear in the conformal blocks of $[JJ]$, while the last two are parity odd and appear with $[\widetilde{JJ}]$. $D^{t}_{R,i}$ is obtained by permuting $(1\leftrightarrow3, 4\leftrightarrow2)$ in $D_{L,i}^{t}$. 

At leading order in $u\ll v \ll 1$, the crossing equation becomes
\begin{align}\label{eqn:CrossingJJJJLO}
C_{J}^{2}\frac{H_{12}H_{34}}{(P_{12})^{d}(P_{34})^{d}}=\sum_{\ell,i,j}\big[&P^{(ij)}_{[JJ]_{0,\ell}}D_{L,i}^{t} D_{R,j}^{t} W_{[JJ]_{0,\ell}}^{t}+P^{(ij)}_{[JJ]_{1,\ell}}D_{L,i}^{t} D_{R,j}^{t}W_{[JJ]_{1,\ell}}^{t} \nonumber \\&+P^{(ij)}_{[\widetilde{JJ}]_{0,\ell}}\tilde{D}_{L,i}^{t} \tilde{D}_{R,j}^{t}W_{\widetilde{[JJ]}_{0,\ell}}^{t}\big].
\end{align} 
As explained in Appendix \ref{App:DegeneracyEquations}, there are two degeneracy conditions among the four point function tensor structures. In practice, it is simpler to match particular dot products appearing in the four point function tensor structures, taking into account the aforementioned degeneracies.

We find a few simplifications when solving (\ref{eqn:CrossingJJJJLO}). First, both the parity-even differential operators and the parity-odd differential operator $\tilde{D}_{L,+}$ are symmetric under the exchange of $1\leftrightarrow4$. Therefore they only appear when $\ell$ is even. $\tilde{D}^{t}_{L,-}$, on the other hand, is odd under this exchange and appears when $\ell$ is odd. Second, for any coefficient matrix $P^{(ij)}$ constructed from (\ref{eqn:DefineP}), the cross terms, $D_{L,1}^{t}D_{R,2}^{t}$ and $D_{L,1}^{t}D_{R,2}^{t}$, give sub-dominant contributions compared to $D_{L,1}^{t}D_{R,1}^{t}$ and $D_{L,2}^{t}D_{R,2}^{t}$, and can be ignored. Finally, we find from identity matching that to leading order in $\ell$, $P^{(22)}_{[JJ]_{0,\ell}}=0$ and $P^{(11)}_{[JJ]_{1,\ell}}=0$. 

To summarize, we find that at leading order, the twist-2 parity even states only contribute through $D^{t}_{L,1}D^{t}_{R,1}$, and the twist-4 ones only contribute through $D^{t}_{L,2}D^{t}_{R,2}$. This is a nice simplification as we do not have to worry about a matrix of OPE coefficients. The differential operators for each double trace state are given by: 
\begin{align}\label{eqn:DDoubleTwistForJJJJ}
&D_{[JJ]_{0,\ell}}^{t}\equiv D_{L,1}^{t}D_{R,1}^{t}\big|_{\Delta=2+\ell,\ell}, \nonumber\\
&D_{[JJ]_{1,\ell}}^{t}\equiv D_{L,2}^{t}D_{R,2}^{t}\big|_{\Delta=4+\ell,\ell},\nonumber\\
&D_{[\widetilde{JJ}]_{0,\ell}}^{t}\equiv \frac{1}{4}\big[(1+(-1)^{\ell})\tilde{D}_{L,+}^{t}+(1+(-1)^{\ell+1})\tilde{D}_{L,-}^{t}\big]\big[(1+(-1)^{\ell})\tilde{D}_{R,+}^{t}+(1+(-1)^{\ell+1})\tilde{D}_{R,-}^{t}\big],
\end{align}
with a corresponding OPE coefficient $P_{\mathcal{O}}$. For the odd differential operators we have grouped the even and odd spin differential operators together. In practice we should separate these contributions, split the sum over even and odd spins, approximate as an integral, and then solve. However, we find that parity odd states of even and odd spin yield contributions of the same form, so we can only determine the sum of their OPE coefficients, which is denoted by $P_{[\widetilde{JJ}]_{0,\ell}}$.

Matching all the dot products that appear in $H_{12}H_{34}$, we find the OPE coefficients of double-twist states at leading order in $1/\ell$ to be
\be\label{eqn:PForJJJJU1}
P_{[JJ]_{0,\ell}}=\frac{\sqrt{\pi}C_{J}^{2}}{2^{2\ell+2}} \ell^{-\frac{7}{2}}, \hspace{0.5cm}
P_{[JJ]_{1,\ell}}=\frac{\sqrt{\pi}C_{J}^{2}}{2^{2\ell+6}} \ell^{-\frac{3}{2}}, \hspace{0.5cm}
P_{[\widetilde{JJ}]_{0,\ell}}=\frac{\sqrt{\pi}C_{J}^{2}}{2^{2\ell+2}} \ell^{-\frac{5}{2}}.
\ee

\subsubsection{$\SU(N)$}

The $\SU(N)$ case is similar to the $\U(1)$ case with extra structures from the global symmetry indices. At leading order in the lightcone limit, the s-channel decomposition is dominated by the identity exchange and the 4-point function factorizes. This contribution is reproduced in the t-channel by large spin double-twist states. As in Section \ref{subsec:JJffIdentitySUN}, two conserved currents can form 6 types of double-twist states corresponding to different representations of $\SU(N)$. For each type, there is a crossing equation similar to (\ref{eqn:CrossingJJJJLO}) that requires 3 families of double-twist operators with different twists as given in Table \ref{table:JJJJDouble-Twists}. 
 
For the parity even operators the same selection rules hold as in the purely scalar case, operators of (odd) even spin appear in representations (anti)symmetric under the exchange of adjoint indices. There are no such selection rules for the parity odd sector and we will need to keep track of extra minus signs relative to the scalar case when operators of odd spin appear in representations symmetric with respect to the adjoint indices and vice versa. This is the origin of the factor $(-1)^{\ell_{\bold{r}}}$ that appears multiplying $P_{\mathcal{O}}^{(ij)}$ in Eq.~(\ref{eqn:SU(N)_rules}). 

Exchange symmetry for the parity odd operators implies we need the following differential operator
\begin{align}
D_{\widetilde{[JJ]},\bold{r},n,\ell}^{t}=\frac{1}{4}&\big[(1+(-1)^{\ell-\ell_{\bold{r}}})\tilde{D}_{L,+}^{t}+(1+(-1)^{\ell-\ell_{\bold{r}}+1})\tilde{D}_{L,-}^{t} \big]\nonumber\\ &\big[(1+(-1)^{\ell-\ell_{\bold{r}}})\tilde{D}_{R,+}^{t}+(1+(-1)^{\ell-\ell_{\bold{r}}+1})\tilde{D}_{R,-}^{t} \big],
\end{align}
where $\ell_{\bold{r}}$ is 0 or 1 for representations that appear in the symmetric or antisymmetric product of adjoints. As in the $\U(1)$ case, we choose to group together operators of even and odd spin for each representation because their contributions have the same form. 

Using the crossing symmetry equations for $\SU(N)$ adjoints, the crossing equation at leading order in $u\ll v\ll 1$ can be solved to find the OPE coefficients of double-twist states at leading order in $1/\ell$:
\begin{align}
\bold{P}_{[JJ]_{0,\ell}}=C_{J}^{2}\frac{\sqrt{\pi}}{2^{4+2\ell}} \ell^{-\frac{7}{2}}\bold{P}, \hspace{0.5cm} 
\bold{P}_{[JJ]_{1,\ell}}=C_{J}^{2}\frac{\sqrt{\pi}}{2^{8+2\ell}} \ell^{-\frac{3}{2}} \bold{P}, \hspace{0.5cm} 
\bold{P}_{[\widetilde{JJ}]_{0,\ell}}=C_{J}^{2}\frac{\sqrt{\pi}}{2^{4+2\ell}} \ell^{-\frac{5}{2}}\bold{P},
\end{align}
\noindent with $\bold{P}=\big(\frac{4}{N^2-1},\frac{2}{N},\frac{2 N}{N^2-4},2,1,1\big)$ giving the result for each double-twist state in different representations under $\SU(N)$, which are $\big(I\,, Adj_a\,, Adj_s\,, (S,\bar{A})_a \oplus (A,\bar{S})_a\,, (A,\bar{A})_s\,, (S,\bar{S})_s\big)$. 
 
\subsection{Stress Tensor and Current Matching}
We now solve the crossing equations at the next-to-leading order. We include in the s-channel the contribution of the exchange of the stress tensor $T$, as well as the conserved currents $J^a$ in the $\SU(N)$ case. These contributions are suppressed by a factor of $\sqrt{u}$ relative to the identity contribution. The $\log(v)$ singularity in the conformal blocks of $T$ and $J$ are reproduced on the right hand side by the anomalous dimensions of double-twist operators, $\gamma_{n,\ell}=\gamma_{n}/\ell$. The power of $\ell$ is determined by matching the extra $\sqrt{u}$ suppression in the s-channel. We will only focus on the $n=0$ and $n=0,1$ case for the parity odd and even double-twist operators, respectively.

\subsubsection{U(1)}
Including the exchange of $T_{\mu\nu}$ and yields the following equations for the anomalous dimensions:
\begin{align}\label{eqn:JJJJU1CrossingNLO}
\frac{1}{4}D_{L,T}D_{R,T}W_{T}^{s}=\sum_{\ell,\mathcal{O}}\gamma_{\mathcal{O}}P_{\mathcal{O}} \partial_{\tau_{\mathcal{O}}} D^{t}_{\mathcal{O}}W_{\mathcal{O}}^{t},
\end{align}
where we have implicitly restricted to terms proportional  to $log(v)$. The sum for $\mathcal{O}$ runs over the 3 families of operators $[JJ]_{0,\ell}$, $[JJ]_{1,\ell}$, and $[\widetilde{JJ}]_{0,\ell}$ as given in Table \ref{table:JJJJDouble-Twists}. The differential operators $D_{\mathcal{O}}$ are given in (\ref{eqn:DDoubleTwistForJJJJ}), $D_{L,T}$ is given in (\ref{eqn:DLT}), and $D_{R,T}$ is obtained by permuting $1\leftrightarrow3$, $2\leftrightarrow4$. The OPE coefficients $P_{\mathcal{O}}$ are solutions to the leading order problem and are given in (\ref{eqn:PForJJJJU1}). 

Solving (\ref{eqn:JJJJU1CrossingNLO}), we obtain the anomalous dimensions $\gamma_{\mathcal{O}_{n,\ell}}=\gamma_{\mathcal{O}_n}/\ell$ of operators in Table {\ref{table:JJJJDouble-Twists}} at leading order in $1/\ell$: 
\begin{align}
&\gamma_{[JJ]_{0}}=-\frac{16(3C_{J}-8\pi\lambda_{JJT})^{2}}{3\pi^{4}C_{T}C_{J}^{2}}, \\
&\gamma_{[JJ]_{1}}=-\frac{64(3C_{J}-16\pi\lambda_{JJT})^{2}}{3\pi^{4}C_{T}C_{J}^{2}},\\
&\gamma_{[\widetilde{JJ}]_{0}}=-\frac{32(3C_{J}-8\pi\lambda_{JJT})(16\pi\lambda_{JJT}-3C_{J})}{3\pi^{4}C_{T}C_{J}^{2}}.\label{eqn:gammaJJOddU1}
\end{align}
We see the parity even double-twist states cannot have positive anomalous dimensions while the parity odd anomalous dimensions are not sign definite. Requiring that they be negative semidefinite yields the conformal collider bounds (\ref{eqn:ColliderBoundsForJJT}). The fact that the negativity conditions of (\ref{eqn:gammaJJOddU1}) agrees with that of (\ref{eqn:gammaJfU1}) provides a non-trivial consistency check for our calculations. The results from the $\<JJJJ\>$ analysis is more general because it does not assume the existence of scalar operators in the spectrum.

\subsubsection{$\SU(N)$}

In the non-Abelian case the s-channel includes, in addition to the stress tensor, a conserved current in the $Adj_{a}$ representation. At leading order in $1/\ell$, the resulting anomalous dimensions of the double-twist operators again take the form $\gamma_{\mathcal{O}_{n,\ell}} \approx \gamma_{\mathcal{O}_n}/\ell$, where
\begin{align}
&\boldsymbol{\gamma}_{[JJ]_{0}}=-\frac{8 (C_{J}-4 \pi  \lambda_{JJJ})^2}{\pi ^4 C_{J}^3}\boldsymbol{\gamma}_{J}-\frac{16 (3 C_{J}-8 \pi  \lambda_{JJT})^2}{3 \pi ^4 C_{T} C_{J}^2}\boldsymbol{\gamma}_{T},\label{eqn:gammaJJE0SUN}\\
&\boldsymbol{\gamma}_{[JJ]_{1}}=-\frac{8(8 \pi  \lambda_{JJJ}-C_{J})^2}{\pi ^4 C_{J}^3}\boldsymbol{\gamma}_{J}-\frac{64 (16 \pi  \lambda_{JJT}-3 C_{J})^2}{3 \pi ^4 C_{T} C_{J}^2}\boldsymbol{\gamma}_{T},\label{eqn:gammaJJE1SUN}\\
&\boldsymbol{\gamma}_{[\widetilde{JJ}]_{0}}=-\frac{8 (C_{J}-4 \pi \lambda_{JJJ})(8 \pi \lambda_{JJJ}-C_{J})}{\pi ^4 C_{J}^3}\boldsymbol{\gamma}_{J}-\frac{32 (3 C_{J}-8 \pi  \lambda_{JJT}) (16 \pi  \lambda_{JJT}-3 C_{J})}{3 \pi ^4 C_{T} C_{J}^2}\boldsymbol{\gamma}_{T}, \label{eqn:gammaJJO0SUN}
\end{align}
with $\boldsymbol{\gamma}_{J}=(2N, N, N, 0, 2, -2)$, $\boldsymbol{\gamma}_{T}=(1,1,1,1,1,1)$. They give the result for double-twist operators in different representations $\big(I\,, Adj_a\,, Adj_s\,, (S,\bar{A})_a \oplus (A,\bar{S})_a\,, (A,\bar{A})_s\,, (S,\bar{S})_s\big)$.

The second terms in (\ref{eqn:gammaJJE0SUN}-\ref{eqn:gammaJJO0SUN}) are corrections due to the stress tensor exchange in the s-channel and correspond to the gravitational binding energies in AdS between well separated 2-particle states. The fact that they are the same for different representations is consistent with the universality of gravity. Once again, we find that these anomalous dimensions are negative, or gravity in AdS is attractive at long distances, if and only if the same conformal collider bounds (\ref{eqn:ColliderBoundsForJJT}) are satisfied. 

The corrections to the dimensions from the current exchange are given in the first terms in (\ref{eqn:gammaJJE0SUN}-\ref{eqn:gammaJJO0SUN}). 
In the dual AdS theory, they correspond to the binding energy from non-Abelian interactions for well separated 2-particle states. 
For the parity even states, we find that the sign of the binding energy only depends on the $\SU(N)$ representation of the double-twist state.
For a given family, it is the most negative for the singlet double-twist state and only positive for the symmetric representation $(S,\bar{S})_s$. This matches with our intuition and also agrees with the signs found in the anomalous dimensions of scalar-scalar 2-particle states~\cite{Li:2015rfa}.
However, for the parity odd state this is not true a priori.
The parity odd $\SU(N)$ singlet 2-particle state has non-positive gauge binding energy if and only if the 3-point function coefficients in $\<JJJ\>$ satisfy the bounds (\ref{eqn:boundsForJJJ}). 
This is also equivalent to demanding the sign of the gauge binding energy to be independent of the parity of the bound state.
As noted in Section \ref{subsec:JfJfNLOSUN}, the bounds (\ref{eqn:boundsForJJJ}) also imply that the charge flux one-point function does not change sign at different angles.

\subsection{Higher Spin Symmetry at Large $N$}

As discussed in the $\<JJ\f\f\>$ case, if $C_J$ does not grow with $N$, then the leading $\ell$ anomalous dimension we computed holds only when $N/\ell\ll1$. Even with the freedom of choosing $\lambda_{JJJ}$, the anomalous dimensions of at least one family of double-twist operators would grow with $N$. The $N\sim\ell$ regime is subtle because the large $\ell$ expansion is not separated from the large $N$ expansion. In this subsection we will focus on the opposite regime of $N\gg\ell$, where we will not be able to establish the presence of the double-twist operators. However, the crossing equations and unitarity imply that if $C_J$ and $\<JJJJ\>$ stay finite in the $N\rightarrow\infty$ limit, then the theory must contain an infinite number of higher spin currents at infinite $N$. We will show this by assuming the higher spin currents do not exist and deriving a contradiction. 

We focus on the first two crossing equations in (\ref{eqn:SUNAdjCrossingEquations}), where we first take $N\rightarrow\infty$ and then go to the lightcone limit. These crossing equations become:
\bea
\left(\frac{u}{v}\right)^{3} G^{t}_{I} &=& G^{s}_{(S,\bar{A})}+G^{s}_{(A,\bar{A})}+G^{s}_{(S,\bar{S})}, \label{eqn:IdlargeN}
\\
\left(\frac{u}{v}\right)^3 G^{t}_{Adj_a}&=&\frac{1}{2}\left(G^{s}_{Adj_{a}}+G^{s}_{Adj_{s}}+G^{s}_{(A,\bar{A})}-G^{s}_{(S,\bar{S})}\right), \label{eqn:AdjlargeN}
\ee
where $G^{s,t}_{\bold{r}}$ denotes the part of the 4-point function corresponding to the representation $\bold{r}$ in either the s- or t-channel. These functions implicitly depend on the polarization vectors. 
Since the absence of higher spin currents is assumed, we only need to consider the exchange of the identity and twist 1 operators of spin $\ell\leq2$ in the s-channel. 
Note that $G^{s}_{I}$, which includes the contribution from identity and $T$ exchange, drops out in this limit. $G^{s}_{Adj_{a}}$ includes the contribution from $J$ exchange. This contribution is non-zero if the OPE coefficients for $\<JJJ\>$ are not suppressed, or equivalently, if $C_{J}$ stays finite as $N\rightarrow\infty$. We do not make any assumptions on the twist 1 operators exchanged in the other representations except for the absence of higher spin currents. 

The contribution from the global symmetry current to $G^{s}_{Adj_{a}}$ contains a $\log(v)$ term at leading order in $u$. In addition, we have shown in \cite{Li:2015rfa} that it is impossible to reproduce such a term when each primary operator in the t-channel contributes with the same sign. 
The same problem shows up here when we consider the terms multiplying the $(Z_{1}\cdot Z_{2}) (Z_{3}\cdot Z_{4})$ and $(Z_{1}\cdot P_{2})(Z_{2}\cdot P_{1})(Z_{3}\cdot P_{4})(Z_{4}\cdot P_{3})$ structures in both equations. 
Furthermore, the $\log(v)$ terms in the s-channel $G^{s}_{\bold{r}}$ multiplying these structures also contribute with the same sign.  In particular, if one tries to cancel the logarithm in the RHS of (\ref{eqn:AdjlargeN}) by allowing for the exchange of twist 1 operators in the $(S,\bar{S})$ representation, then this exchange will induce a $\log(v)$ term on the RHS of (\ref{eqn:IdlargeN}), leading to another contradiction. The analysis then reduces to that in \cite{Li:2015rfa}: the exchange of a finite number of higher spin currents in the s-channel cannot remove the logarithms, rather we need to sum over an infinite tower of higher spin currents.  These higher spin currents necessarily transform non-trivially under the global symmetry group of the CFT. 

Let us now consider the exchange of scalars with $\frac{1}{2}<\Delta_{\f}<1$. In fact, it is easy to see that they cannot appear in the s-channel with $\mathcal{O}(1)$ coefficients at $N\rightarrow\infty$ in a unitary CFT. Such scalars would contribute a $\log(v)$ term that could not be cancelled by a sum over higher spin operators. The reason is that, if such higher spin operators existed, they would necessarily violate the unitarity bound. Thus, there are no finite contributions from scalars with $1/2<\Delta_{\f}<1$ to $G^{s}_{\boldsymbol{r}\neq I}$ at $N\rightarrow\infty$.

The argument presented here for the existence of higher spin currents at $N \rightarrow \infty$ (assuming $C_J$ does not grow with $N$) is more general than the one made in \cite{Li:2015rfa}, because we did not need to assume the existence of scalar operators in the spectrum.

\section{Mixed Stress Tensor-Scalar 4-point Functions}
\label{sec:TTOO}
In this section, we study the correlation functions of the form $\<TT\f\f\>$ where $T_{\mu\nu}$ is the stress-energy tensor and $\f$ is a scalar operator of arbitrary dimension.  Since $T_{\mu\nu}$ is conserved it saturates the unitarity bound and has dimension $d$. At leading order in $u\ll1$, the 4-point function will be dominated by the identity contribution in the s-channel. We will assume that the next leading order correction comes from stress-energy tensor exchange. Note that the correlator $\<TTJ\>$ vanishes in three dimensional CFTs~\cite{Giombi:2011rz}. We will also not consider the corrections due to the exchange of a light scalar.

\subsection{Identity Matching}
At leading order in $u$, the 4-point function is approximately given by the factorized form found in generalized free theories. In the t-channel this is reproduced by two families of double-twist operators with even or odd parity, of the schematic form:
\be
[T\f]_{n,\ell}=T_{\mu\nu}(\partial^{2})^{n}\partial_{\sigma_{1}}...\partial_{\sigma_{\ell-2}}\f,\hspace{1cm}\widetilde{[T\f]}_{n,\ell}=\epsilon_{\alpha}^{\ \nu\kappa}T_{\mu\nu}(\partial^{2})^{n}\partial_{\kappa}\partial_{\sigma_{1}}...\partial_{\sigma_{\ell-2}}\f.
\ee
The parity-even operators $[T\phi]_{n,\ell}$ have twist $1+2n+\Delta_\phi$, while the parity-odd operators $[\widetilde{T\phi}]_{n,\ell}$ have twist $2+2n+\Delta_\phi$. The crossing equation at leading order in $u$ is given by
\be
C_{T}\frac{H_{12}^{2}}{(P_{12})^{3}(P_{34})^{\Delta_{\f}}}=& \sum_{n,\ell} P_{[T\f]_{n,\ell}}D^{t}_{[T\f]_{n,\ell}}W_{[T\f]_{n,\ell}}^{t}+P_{\widetilde{[T\f]}_{n,\ell}}D^{t}_{\widetilde{[T\f]}_{n,\ell}}W_{\widetilde{[T\f]}_{n,\ell}}^{t}.
\ee
We solve this equation at leading order in $v$, which restricts the t-channel operators to have $n=0$. The t-channel differential operators are constructed in Appendix \ref{App:TTffDifferentialOperators}, and take the form
\begin{align}
D^{t}_{[T\f]_{0,\ell}}=&\bigg(\frac{6}{(\Delta_{\f}+\ell-2)(\Delta_{\f}+\ell-1)}(D^{t}_{11})^{2}\Sigma_{L}^{2,0}-\frac{4}{\Delta_{\f}+\ell-1} D^{t}_{14}D^{t}_{11}\Sigma_{L}^{1,1}+(D^{t}_{14})^{2}\Sigma_{L}^{0,2}\bigg)\nonumber
\\&\bigg(\frac{6}{(\Delta_{\f}+\ell-2)(\Delta_{\f}+\ell-1)}(D^{t}_{22})^{2}\Sigma_{R}^{2,0}-\frac{4}{\Delta_{\f}+\ell-1} D^{t}_{23}D^{t}_{22}\Sigma_{L}^{1,1}+(D^{t}_{23})^{2}\Sigma_{R}^{0,2} \bigg),
\\
D^{t}_{\widetilde{[T\f]}_{0,\ell}}=&\bigg(-\frac{3}{\Delta_{\f}+\ell-1}D^{t}_{11}\tilde{D}^{t}_{1}\Sigma_{L}^{1,0}+D^{t}_{14}\tilde{D}^{t}_{1}\Sigma_{L}^{0,1}\bigg)\bigg(-\frac{3}{\Delta_{\f}+\ell-1}D^{t}_{22}\tilde{D}^{t}_{1}\Sigma_{R}^{1,0}+D^{t}_{23}\tilde{D}^{t}_{1}\Sigma_{R}^{0,1}\bigg).
\end{align}
We match all dot products appearing in $H_{12}^{2}$ in a basis of independent tensor structures. In particular, we matched the coefficient of three structures: $(Z_{1}\cdot P_{2})^2(Z_{2}\cdot P_{1})^{2}$, $(Z_{1}\cdot Z_{2})^{2}$, and $(Z_{1}\cdot Z_{2})(Z_{1}\cdot P_{2})(Z_{2}\cdot P_{1})$. This results in a linearly dependent set of equations that can be solved for the OPE coefficient at leading order in $1/\ell$:
\bea
P_{[T\f]_{0,\ell}}\approx C_{T}\frac{\sqrt{\pi } 2^{-\Delta_{\f}}}{2^{2\ell}3 \Gamma (\Delta_{\f})}\ell^{\Delta_{\f}-\frac{1}{2}}, \hspace{1cm}P_{\widetilde{[T\f]}_{0,\ell}}\approx C_{T}\frac{\sqrt{\pi } 2^{-\Delta_{\f}}}{2^{2\ell}3 \Gamma (\Delta_{\f})}\ell^{\Delta_{\f}-\frac{7}{2}}.
\eea

\subsection{Stress Tensor Matching}
At the next leading order in $u$ we include the exchange of $T$ in the s-channel. This contribution is suppressed by a factor of $\sqrt{u}$ compared to the identity.
This implies that the anomalous dimensions are $1/\ell$ suppressed at large $\ell$. At leading order in $v$ and next-to-leading order in $u$, the crossing equation takes the form 
\begin{align}\label{eqn:TTffCrossingNLO}
\frac{\lambda_{\f\f T}}{4\sqrt{C_{T}}}\mathcal{D}_{T}W_{T}^{s}=&\sum_{\ell}\gamma_{[T\f]_{0,\ell}}P_{[T\f]_{0,\ell}}\partial_{\tau_0}D^{t}_{[T\f]_{0,\ell}}W_{[T\f]_{0,\ell}}^{t}+
\gamma_{\widetilde{[T\f]}_{0,\ell}} P_{\widetilde{[T\f]}_{0,\ell}}\partial_{\tilde{\tau}_0}D^{t}_{\widetilde{[T\f]}_{0,\ell}}W_{\widetilde{[T\f]}_{0,\ell}}^{t}, 
\end{align}
where we project onto the $\log(v)$ terms. The differential operator $\mathcal{D}_{T}$ is constructed in Appendix \ref{App:TTffDifferentialOperators} by matching to the 3-point function of the stress tensor. This 3-point function depends on two coefficients, $C_T$ and $\lambda_{TTT}$, where $C_T$ is the central charge and $\lambda_{TTT}$ is defined explicitly in terms of the 3-point function structures in Appendix \ref{App:TTffDifferentialOperators}.\footnote{It is related to the coefficient $t_4$ used in~\cite{Buchel:2009sk} by the relation $\lambda_{TTT} = \frac{3 C_T (60-t_4)}{2^{10}\pi}$.}  Solving $(\ref{eqn:TTffCrossingNLO})$ then gives the anomalous dimensions of the leading double-twist states $\gamma_{\mathcal{O}_{n,\ell}}=\gamma_{\mathcal{O}_{0}}/\ell$ with coefficients
\bea
\gamma_{[T\f]_{0}}&=&-\frac{16 \Delta_{\f} (3 C_{T}-16 \pi  \lambda_{TTT}) \Gamma (\Delta_{\f})}{\pi ^{7/2} C_{T}^2 \Gamma \left(\Delta_{\f}-\frac{1}{2}\right)},
\\
\gamma_{[\widetilde{T\f}]_{0}}&=&-\frac{8 \Delta_{\f} (128 \pi  \lambda_{TTT}-21 C_{T}) \Gamma (\Delta_{\f})}{\pi ^{7/2} C_{T}^2 \Gamma \left(\Delta_{\f}-\frac{1}{2}\right)}.
\eea
In an AdS bulk description, these anomalous dimensions corresponds to the correction to the energy of well separated graviton-scalar two particle states from gravitational interactions. We expect gravity to be attractive at large distances and requiring the anomalous dimensions to be negative semidefinite yields the following constraints
\bea\label{eqn:TTTBound}
C_{T}>0, \quad \frac{21C_{T}}{128\pi} \leq\lambda_{TTT}\leq \frac{3C_{T}}{16\pi} .
\eea
In a theory of free bosons and fermions in three dimensions we have \cite{Osborn:1993cr}
\begin{align}
C_{T}=\frac{3 (2 n_{\psi}+n_{\f})}{32 \pi ^2}, \hspace{1cm} \lambda_{TTT}=\frac{9 (16 n_{\psi}+7 n_{\f})}{4096 \pi ^3}.
\end{align}
It follows that free theories saturate the bounds:
\begin{align}
\frac{\lambda_{TTT}}{C_{T}}\bigg|_{n_{\f}=0}=\frac{3}{16\pi}, \qquad \frac{\lambda_{TTT}}{C_{T}}\bigg|_{n_{\psi}=0}=\frac{21}{128\pi}.
\end{align}
Moreover, the bounds on $\lambda_{TTT}$ above correspond to $n_{\f}\geq0 \ \text{and} \ n_{\psi}\geq0$, which match the conformal collider bounds found in \cite{Buchel:2009sk}, Eq.~(3.43).

\section{Superconformal Field Theories}
\label{sec:SCFT}

It can be shown that every 3d superconformal field theory (SCFT) trivially satisfies the conformal collider bounds on $\lambda_{JJT}$, $\lambda_{TTT}$, and the conjectured bound on $\lambda_{JJJ}$. For the moment we will assume the conserved current corresponds to a flavor (non-R) symmetry. Working in $\mathcal{N}=1$ superspace, it was found in \cite{Nizami:2013tpa} that the parity even 3-point functions of conserved operators is fixed up to an overall coefficient.\footnote{In \cite{Nizami:2013tpa} any multiplet containing conserved operators is referred to as a supercurrent, but we will follow the terminology of \cite{Buchbinder:2015wia,Buchbinder:2015qsa} and refer to only the supermultiplet containing the stress-energy tensor as the supercurrent.}

Since these correlation functions are fixed up to an overall coefficient we can calculate $\lambda_{JJJ}$, $\lambda_{JJT}$, and $\lambda_{TTT}$ in free supersymmetric theories in terms of the central charges using Section 5 of \cite{Osborn:1993cr}. For general free field theories in three dimensions we have
\begin{align}
\frac{\lambda_{JJJ}}{C_{J}}&=\frac{\sum_{i}2C(\bold{r}_{i})_{mf}+C(\bold{r}_{i})_{s}}{\sum_{i}8\pi(C(\bold{r}_{i})_{mf}+C(\bold{r}_{i})_{s})}\label{eqn:b/CJSUSY},\\
\frac{\lambda_{JJT}}{C_{J}}&=\frac{\sum_{i}3 (2 C(\bold{r}_{i})_{mf}+C(\bold{r}_{i})_{s})}{\sum_{i}16 \pi  (C(\bold{r}_{i})_{mf}+C(\bold{r}_{i})_{s})},\\
\frac{\lambda_{TTT}}{C_{T}}&=\frac{3 (8 n_{mf}+7 n_{s})}{128 \pi  (n_{mf}+n_{s})}.
\end{align}
Here $C(\bold{r}_{i})$ is the index of the representation and the subscripts $mf$ and $s$ stand for Majorana fermions and real scalars respectively. Finally, $n_{mf}$ and $n_{s}$ give the total number of real Majorana fermions and real scalars. In a free supersymmetric theory there are an equal number of Majorana fermions and real scalars, in total and in a given representation of the flavor symmetry, so we have
\begin{align}\label{eqn:3dSUSY3PFs}
\frac{\lambda_{JJJ}}{C_{J}}=\frac{3}{16\pi}, \ \ \ \frac{\lambda_{JJT}}{C_{J}}=\frac{9}{32\pi}, \ \ \ \frac{\lambda_{TTT}}{C_{T}}=\frac{45}{256\pi}.
\end{align}
Although we computed these using the free theory, these results holds for any 3d superconformal field theory.
We also see that $\lambda_{JJT}/C_J$ and $\lambda_{TTT}/C_T$ satisfy the conformal collider bounds on $\<JJT\>$ (\ref{eqn:ColliderBoundsForJJT}) and $\<TTT\>$ (\ref{eqn:TTTBound}). 
These values lead to a uniform integrated, energy distribution measured at spatial infinity after a local perturbation is created by a conserved current (\cite{Chowdhury:2012km}, Eq. (6.14)) or the stress-energy tensor (\cite{Buchel:2009sk}, Eqs. (3.8) and (3.43)). 
We also note that the ratio $\lambda_{JJJ}/C_J$ satisfies the conjectured bounds on $\<JJJ\>$ in (\ref{eqn:boundsForJJJ}). 
We have computed the charge correlator in Appendix \ref{App:Charge1PF} in terms of $\lambda_{JJJ}/C_J$, where we find that the charge flux is also uniform for the supersymmetric value given above. 

We will now move on to the case of R-symmetry currents. We will start with $\mathcal{N}=2$ supersymmetric theories which have a $\U(1)_{R}$ symmetry. For clarity we can make the replacement $C(\bold{r}_{i})_{mf,s}\rightarrow (q_{i}^{mf,s})^{2}$, where $q_{i}$ denotes the charge under the $\U(1)_{R}$ symmetry. As shown in \cite{Buchbinder:2015qsa}, the three point function of the supercurrent is fixed up to an overall constant, so once again we can calculate $\lambda_{JJT}/C_J$ and $\lambda_{TTT}/C_T$ using a free field theory of chiral multiplets ($\lambda_{JJJ}$ does not appear since we have three $\U(1)$ currents). A free $\mathcal{N}=2$ chiral multiplet consists of a scalar with R-charge 1/2 and a fermion of R-charge -1/2, so the results for $\lambda_{JJT}$ and $\lambda_{TTT}$ found in (\ref{eqn:3dSUSY3PFs}) still hold. 

Next we consider theories with $\mathcal{N}=3$ SUSY, which have an $\SO(3)$ R-symmetry. Once again we have a one parameter family of free field theories, this time with an equal number of complex scalars and fermions in the spinor representation of the R-symmetry group \cite{Nizami:2013tpa}. The three point function of the supercurrent $\mathcal{J}_{\alpha}$ is fixed up to an overall constant \cite{Buchbinder:2015qsa}, so we will find the same ratios as before.

Finally, we will study theories with $\mathcal{N}=4$ SUSY, which were extensively studied in \cite{Buchbinder:2015wia}. What makes these theories special is that the R-symmetry group, $\SO(4)$, is locally isomorphic to $\SU(2)_{L}\times\SU(2)_{R}$. Therefore, we can consider a two-parameter family of free field theories, consisting of hypermultiplets $q^{i}$ and $q^{\tilde{i}}$ in the (2,1) and (1,2) representation of the R-symmetry group, respectively. Here we label representations by their dimension and note that each hypermultiplet consists of four real scalars and four Majorana fermions. The supercurrent is given by a real scalar superfield $\mathcal{J}$ and its three point function has two linearly independent tensor structures parametrized by $d_{\mathcal{N}=4}$ and $\tilde{d}_{\mathcal{N}=4}$. This is in contrast to theories with less supersymmetry, where the three point function is fixed up to an overall constant. The resolution found in \cite{Buchbinder:2015wia} was that the $\mathcal{N}=4$ supermultiplet contains two $\mathcal{N}=3$ supermultiplets,  $\mathcal{S}$ and $\mathcal{J}_{\alpha}$, the latter being the $\mathcal{N}=3$ supercurrent, and that only one of the tensor structures contributes to $\<\mathcal{J}_{\alpha}\mathcal{J}_{\beta}\mathcal{J}_{\gamma}\>$.

We can now repeat the above analysis for a free theory with $m$ left hypermultiplets and $n$ right hypermultiplets, specializing to study the $\SU(2)_{L}$ R-current, $J_{\mu,L}$. The correlation functions $\<J_{L,\mu}J_{L,\nu}T_{\rho\sigma}\>$  and $\<J_{L,\mu}J_{L,\nu}J_{L,\rho}\>$ are once again fixed up to an overall coefficient and our results for $\lambda_{JJJ}$ and $\lambda_{JJT}$ hold as before with $C_{J}\rightarrow C_{J,L}$. The analogous substitution will also have to be made for the  $\SU(2)_{R}$ current.

Using these results, we can determine the large spin spectrum of double-twist states involving an R-symmetry current. We will consider only the exchange of the R-current itself and the stress-energy tensor in the s-channel and not the exchange of light scalars or other conserved currents. For $\mathcal{N}=2$, $\<J_R J_R J_R\>$ vanishes so all double-twist states formed from two R-currents or a R-current and a scalar will have negative anomalous dimensions. 

Moving on to theories with $\mathcal{N}=3$ SUSY, we first need to find the ratio $C_{T}/C_{J}$. The R-current and stress-energy tensor lie in the same supermultiplet, so we can calculate this ratio either by expanding the supercurrent two point function as in~\cite{Li:2014gpa} or by calculating it in a free theory. Using Eqs.~(5.5) and (5.6) in \cite{Osborn:1993cr}, we find that $C_{T}/C_{J}=3$. We then find that the coefficient of anomalous dimensions for double-twist states formed from the R-current and a scalar in the adjoint representation of the R-symmetry group become
\begin{align}
&\gamma_{[J_R \f]_0}=\left(-\frac{3 (\Delta_{\f} +2) \Gamma (\Delta_{\f} )}{\pi ^{7/2} C_{T} \Gamma \left(\Delta_{\f} -\frac{1}{2}\right)},-\frac{3 (\Delta_{\f} +1) \Gamma (\Delta_{\f} )}{\pi ^{7/2} C_{T} \Gamma \left(\Delta_{\f} -\frac{1}{2}\right)},-\frac{3 (\Delta_{\f} -1) \Gamma (\Delta_{\f} )}{\pi ^{7/2} C_{T} \Gamma \left(\Delta_{\f} -\frac{1}{2}\right)}\right),
\\
&\gamma_{[\widetilde{J_R \f}]_0}=\left(-\frac{12(\Delta_{\f}+1)\Gamma(\Delta_{\f})}{\pi^{7/2}C_{T}\Gamma\left(\Delta_{\f}-\frac{1}{2}\right)},-\frac{6(2\Delta_{\f}+1)\Gamma(\Delta_{\f})}{\pi^{7/2}C_{T}\Gamma(\Delta_{\f}-\frac{1}{2})},\frac{6(1-2\Delta_{\f})\Gamma(\Delta_{\f})}{\pi^{7/2}C_{T}\Gamma(\Delta_{\f}-\frac{1}{2})}\right),
\end{align}
where we have used that $\SO(3)\simeq\SU(2)$ and expressed our results in the basis $\big(I\,, Adj_a,(S,\bar{S})\big)$. Note that all the anomalous dimensions are non-positive if $\Delta_{\f}\geq1$. This bound is nothing more than the unitarity bound for 3d scalars in the adjoint representation of the $\SO(3)$ R-symmetry group \cite{Minwalla:1997ka}.  Furthermore, scalars which saturate this bound belong to a short representation of the superconformal algebra and their leading anomalous dimension asymptotic for the parity even $(S,\bar{S})$ double-twist state vanishes. Finally, for double-twist states formed from two R-currents, we have
\begin{align}
\gamma_{[J_RJ_R]_0}&=\left(-\frac{9}{\pi ^4 C_{T}},-\frac{6}{\pi ^4 C_{T}},0\right),
\\
\gamma_{[J_RJ_R]_1}&=\left(-\frac{72}{\pi ^4 C_{T}},-\frac{60}{\pi ^4 C_{T}},-\frac{36}{\pi ^4 C_{T}}\right),
\\
\gamma_{[\widetilde{J_RJ_R}]_0}&=\left(-\frac{24}{\pi ^4 C_{T}},-\frac{18}{\pi ^4 C_{T}},-\frac{6}{\pi ^4 C_{T}}\right).
\end{align}
Once again, all the anomalous dimensions either vanish or are negative.

Finally, we will consider theories with $\mathcal{N}=4$ SUSY and focus on the $\SU(2)_{L}$ R-current. For the moment we only consider the effect of the R-currents and $T_{\mu\nu}$ in the s-channel. The supercurrent multiplet also contains a dimension 1 scalar which will contribute to the anomalous dimensions at the same order which we will consider later. 

If we consider a double-twist state formed from the $\SU(2)_{L}$ R-current and a scalar in the same representation (i.e., adjoint of $\SU(2)_{L}$ and singlet of $\SU(2)_{R}$), we find that the contribution of the R-current and stress energy tensor to the anomalous dimension asymptotics is given by
\bea
\gamma_{[{J_{L}\f}]}=\left(\frac{\Gamma (\Delta_{\f} ) (-2 C_{T}-3 C_{J,L} \Delta_{\f} )}{\pi ^{7/2} C_{T} C_{J,L} \Gamma \left(\Delta_{\f} -\frac{1}{2}\right)},-\frac{\Gamma (\Delta_{\f} ) (C_{T}+3 C_{J,L} \Delta_{\f} )}{\pi ^{7/2} C_{T} C_{J,L} \Gamma \left(\Delta_{\f} -\frac{1}{2}\right)},\frac{\Gamma (\Delta_{\f} ) (C_{T}-3 C_{J,L} \Delta_{\f} )}{\pi ^{7/2} C_{T} C_{J,L} \Gamma \left(\Delta_{\f} -\frac{1}{2}\right)}\right),  \ \ \ \\
\gamma_{\widetilde{[{J_{L}\f}]}}=\left(-\frac{4 \Gamma (\Delta_{\f} ) (C_{T}+3 C_{J,L} \Delta_{\f} )}{\pi ^{7/2} C_{T} C_{J,L} \Gamma \left(\Delta_{\f} -\frac{1}{2}\right)},-\frac{2 \Gamma (\Delta_{\f} ) (C_{T}+6 C_{J,L} \Delta_{\f} )}{\pi ^{7/2} C_{T} C_{J,L} \Gamma \left(\Delta_{\f} -\frac{1}{2}\right)},\frac{2 \Gamma (\Delta_{\f} ) (C_{T}-6 C_{J,L} \Delta_{\f} )}{\pi ^{7/2} C_{T} C_{J,L} \Gamma \left(\Delta_{\f} -\frac{1}{2}\right)}\right). \ \ \
\eea
So the contribution is non-positive for all double trace states if and only if $\Delta_{\f}\geq \frac{C_{T}}{3C_{J,L}}$. When this inequality is saturated the parity even $(S,\bar{S})$ term vanishes to leading order. 

For the double-twist states formed from two $\SU(2)_{L}$ R-currents, the contribution of $J_{\mu,L}$ and $T_{\mu\nu}$ to the anomalous dimensions becomes
\bea
\gamma_{[J_{L}J_{L}]_{0}}=\left(-\frac{2 C_{T}+3 C_{J,L}}{\pi ^4 C_{T} C_{J,L}},-\frac{C_{T}+3 C_{J,L}}{\pi ^4 C_{T} C_{J,L}},\frac{C_{T}-3 C_{J,L}}{\pi ^4 C_{T} C_{J,L}}\right),
\\
\gamma_{[J_{L}J_{L}]_{1}}=\left(-\frac{8 (C_{T}+6 C_{J,L})}{\pi ^4 C_{T} C_{J,L}},-\frac{4 (C_{T}+12 C_{J,L})}{\pi ^4 C_{T} C_{J,L}},\frac{4 (C_{T}-12 C_{J,L})}{\pi ^4 C_{T} C_{J,L}}\right),
\\
\gamma_{\widetilde{[J_{L}J_{L}]}_{0}}=\left(-\frac{4 (C_{T}+3 C_{J,L})}{\pi ^4 C_{T} C_{J,L}},-\frac{2 (C_{T}+6 C_{J,L})}{\pi ^4 C_{T} C_{J,L}},\frac{2 (C_{T}-6 C_{J,L})}{\pi ^4 C_{T} C_{J,L}}\right).
\eea
The above quantities are all negative if $C_{T}\leq3C_{J,L}$. When this inequality is saturated the contribution of $T_{\mu\nu}$ and $J_{\mu,L}$ vanishes for twist two, parity even double-twist states in the $(S,\bar{S})$ representation. Similar results can be found for the $\SU(2)_{R}$ current by letting $C_{J,L}\leftrightarrow C_{J,R}$. 

If these inequalities are not satisfied then for some double trace states the contribution of the R-current is greater than the contribution from the stress energy tensor. This may be related to a non-Abelian version of the weak gravity conjecture \cite{ArkaniHamed:2006dz}.

To be complete we would also have to include the dimension 1 scalar superconformal primary of the supercurrent multiplet, $J$. One might wonder whether, after taking it into account, we will obtain a convex spectrum once the relevant unitarity bounds are satisfied. There is a simple way to see this cannot be the case. As mentioned earlier, upon reduction from $\mathcal{N}=4$ superspace to $\mathcal{N}=3$ superspace, the supercurrent $\mathcal{J}$ splits into two superfields, $\mathcal{J}_{\alpha}$, which contains the $\mathcal{N}=3$ R-symmetry currents, and $\mathcal{S}$ which contains the dimension 1 scalar and the missing $\mathcal{N}=4$ R-currents. In \cite{Buchbinder:2015wia} they found that both $\<\mathcal{J_{\a}}\mathcal{J_{\b}}\mathcal{S}\>$ and $\<\mathcal{S}\mathcal{S}\mathcal{S}\>$ are determined by a single parameter $\tilde{d}_{\mathcal{N}=4}$, which in a free theory is proportional to $m-n$. Therefore if the theory has $m=n$, or $\tilde{d}_{\mathcal{N}=4}$=0, the scalar makes no contribution and some of the double trace states will still have a concave spectrum in twist space at large spin.

The effect of the dimension 1 scalar on the double twist states of two $J_{L}$ currents is to shift the anomalous dimension of only the parity even, twist two state by
\bea
\delta\gamma_{[JJ]_{0}}=-\frac{4 (C_{T}-6 C_{J,L})^2}{3 \pi ^4 C_{T} C_{J,L}^2}\big(1,1,1\big).
\eea
One question that arises is how to to reproduce the $\mathcal{N}=3$ results, where convexity was automatically satisfied, from our $\mathcal{N}=4$ results. The natural choice is to identify the $\mathcal{N}=3$ R-symmetry currents as the generators of the diagonal subgroup  $\SU(2)_{L}\times\SU(2)_{R}\big{|}_{diag}\simeq \SU(2)$, given by $J_{d}^{a}=J^{a}_{L,\mu}+J^{a}_{R,\mu}$. We have re-introduced the adjoint indices for the currents to emphasize we are considering the diagonal subgroup. The analysis then exactly mimics the $\mathcal{N}=3$ case, assuming that when studying $\<JJ\f\f\>$ the scalar also transforms in the adjoint representation of the diagonal subgroup.

\section{Discussion}
\label{sec:discussion}

By studying the conformal bootstrap equations in the lightcone limit, we have generalized the CFT argument for the cluster decomposition principle to operators with spin. In doing so we have derived the existence of large spin double-twist conformal primaries constructed from spinning operators. We computed their anomalous dimensions and showed that they turn off as $\ell\rightarrow\infty$. In an AdS dual description, a large spin double-twist operator corresponds to a two particle state with a separation $\sim \log (\ell)$. The anomalous dimension then describes the binding energy induced by the exchange of light particles such as gauge bosons and gravitons. 

In this work we discovered a connection between the signs of the anomalous dimensions and the conformal collider bounds (\ref{eqn:ColliderBoundsForJJT}) and (\ref{eqn:TTTBound}) in parity-symmetric 3d CFTs. In all cases under consideration, the anomalous dimensions due to stress-energy tensor exchange are negative semi-definite if and only if the conformal collider bounds are satisfied. 
These anomalous dimensions are expected to be non-positive from the bulk point of view, since we expect gravity to be attractive at large distances. We can turn the logic around and conclude that the conformal collider bounds, combined with our analysis, provide a pure CFT argument for the attractiveness of bulk gravity at long distances, which does not require a large $N$ limit and holds for all unitary theories.

It would be interesting to see if the same bounds can be derived from more basic axioms such as unitarity or causality. The connection to unitarity and deep inelastic scattering (DIS) arguments were explored in \cite{Komargodski:2012ek,Kulaxizi:Presentation}. In \cite{Kulaxizi:2010jt} it was also seen that the conformal collider bounds on $\langle TTT\rangle$ in 4d can be derived from unitarity if the stress tensor is the only spin-2 conserved operator in the $TT$ OPE that can get a vacuum expectation value at finite temperature. 

For classes of large $N$ CFTs it has been shown that causality is related to energy flux positivity \cite{Buchel:2009sk,Hofman:2009ug,Camanho:2009vw,Brigante:2008gz,Brigante:2007nu,Buchel:2009tt,deBoer:2009pn,deBoer:2009gx}. Causality of bulk gravity is also related to the negativity of the anomalous dimensions due to the exchange of $T_{\mu\nu}$ in the direct channel. The anomalous dimensions of double-twist states of large spin and twist formed from scalars in large $N$ CFTs were found to be related to Shapiro time delay in the bulk \cite{Cornalba:2006xm,Cornalba:2006xk,Cornalba:2007zb}. This result was generalized to arbitrary double-twist states formed from scalars in large $N$ CFTs \cite{Camanho:2014apa}. At least in large $N$ theories, there is an intimate relation between the negativity of the anomalous dimensions or the attractiveness of gravity at long distances, causality in the bulk, and positivity of integrated energy one point functions in the Lorentzian CFT. Our work provides the direct link between the anomalous dimensions and the energy positivity conditions, and also extends the discussion beyond the large $N$ limit to generic, non-perturbative CFTs. In this regime, it may also be possible to establish the connection to causality along the lines of~\cite{Hartman:2015lfa}. 

Furthermore, we speculated that a new conformal collider-like bound (\ref{eqn:boundsForJJJ}) may exist for $\<JJJ\>$, that its undetermined coefficient must lie in between the free fermion and free boson values. This is equivalent to the requirement that the signs of the anomalous dimensions due to $J$ exchange only depend on the global symmetry representations of the double-twist states and not on their spin or parity. In the conformal collider set-up, we computed the charge 1-point function in terms of $\<JJJ\>$. We find that the same bound implies that the expectation value of the integrated charge flux is positive at all angles after a positive amount of charge is injected with a local perturbation created by $J$, thus putting constraints on the charge flux asymmetry. In all cases we are aware of in three dimensions, this bound holds and it would be interesting to see if there exists a proof or explicit counterexamples. 

We also applied our results to the study of 3d superconformal field theories. The conformal collider bounds and conjectural bound for $\<JJJ\>$ are found to be satisfied for a theory with any amount of SUSY. The value of the corresponding 3-point functions result in uniform energy/charge flux distributions at infinity after a local perturbation. In addition, we find that for SCFTs with $\mathcal{N}=2,3$ SUSY the exchange of the supercurrent multiplet induces non-positive anomalous dimensions for several families of double-twist operators formed by two R-currents or one R-current and one scalar in the adjoint of the R-symmetry. 
This does not seem to hold for theories with $\mathcal{N}=4$ symmetry. The distinguishing characteristic of $\mathcal{N}=4$ theories in comparison to CFTs with less SUSY in three dimensions is that the R-symmetry group is locally isomorphic to a product of groups.

We have restricted ourselves to 3d CFTs since this is the only case where all the conformal blocks are currently known. Given recent progress in calculating conformal blocks for $\<JJ\f\f\>$ \cite{Rejon-Barrera:2015bpa}, it should be straightforward to extend the arguments for $\<JJT\>$ to higher dimensions. Generalizing our study of $\<TT\f\f\>$ and $\<JJJJ\>$ to 4d will require more work. Another straightforward generalization will be to include the effects of parity-violating couplings in 3d. We have also not yet studied $\<TTTT\>$ in three dimensions as incorporating all possible degeneracy equations requires an intricate analysis~\cite{Dymarsky:2013wla} and it is not required to probe the conformal collider bounds.

Our work is just a first step in analytically solving the bootstrap equations for spinning operators. Some simple extensions would be to include external fermions, operators of higher spin, or non-conserved spin 1 and 2 operators. By studying anomalous dimensions of double-twist states with twist comparable to or much greater than their spin we can also hope to derive the more stringent bounds of \cite{Camanho:2014apa} on corrections to Einstein gravity in AdS$_{4}$. 

Finally, we should note these correlation functions have not yet been studied with the numerical bootstrap. It will be exciting to see if these bounds can be derived there. Studying the 4-point functions of these conserved operators, both analytically and numerically, is a key step in mapping out the space of consistent CFTs.

\section*{Acknowledgements}
We thank Tom Hartman, Jared Kaplan, Zuhair Khandker, Filip Kos, Petr Kravchuk, Juan Maldacena, David Simmons-Duffin, Fernando Rejon-Barrera, Andreas Stergiou, Junpu Wang, Matthew Walters and Sasha Zhiboedov for discussions. This work is supported by NSF grant 1350180.  DP and DL thank the Aspen Center for Physics for its hospitality during the completion of this work, supported by NSF Grant 1066293. DL thanks the Simons Center for Geometry and Physics at Stony Brook University for its hospitality during the completion of this work. DP receives additional support as a Martin. A and Helen Chooljian Founders' Circle Member at IAS. DL and DM also thank the Institute for Advanced Study for its hospitality during the completion of this work.

\appendix
\section{Collinear Conformal Blocks at Large Spin}
\label{sec:Appendix_A}
In the small $v$ limit the t-channel conformal blocks become \cite{DO3}
\bea
g^{\{\Delta_{i}\}}_{\tau,\ell}(v,u)=v^{\frac{1}{2}\tau}(1-u)^{\ell}\ _{2}F_{1}\bigg(\frac{1}{2}(\tau+2\ell)+a,\frac{1}{2}(\tau+2\ell)+b;\tau+2\ell;1-u\bigg),
\eea
where we use $\{\tau, \ell\}$ in place of $\{\Delta,\ell\}$. Here $a=\frac{1}{2}(\Delta_{4}-\Delta_{1})$ and $b=\frac{1}{2}(\Delta_{3}-\Delta_{2})$. The form is same for the s-channel blocks in  small $u$ limit but with $u \rightarrow v$, $a=\frac{1}{2}(\Delta_{2}-\Delta_{1})$, and $b=\frac{1}{2}(\Delta_{3}-\Delta_{4})$. This approximation is sufficient in the s-channel since there we have a finite number of blocks, but we will need to make further approximations in the t-channel.

Using the integral representation of the hypergeometric function, we rewrite this as
\bea
g^{\{\Delta_{i}\}}_{\tau,\ell}(v,u)=v^{\frac{1}{2}\tau}(1-u)^{\ell}\int_{0}^{1}\mathrm{dt}\frac{\Gamma (2 \ell+\tau ) (1-t)^{-b+\ell+\frac{\tau }{2}-1} t^{b+\ell+\frac{\tau }{2}-1} (t (u-1)+1)^{-a-\ell-\frac{\tau }{2}}}{\Gamma (-b+\ell+\frac{\tau }{2}) \Gamma (b+\ell+\frac{\tau }{2})}.\hspace{0.3cm}
\eea
We want to expand the above expression at large $\ell$, where we keep $y\equiv u\ell^{2} \lesssim \mathcal{O}(1)$. Defining $s\equiv\frac{ty}{\ell(1-t)}$ and expanding in this limit yields
\begin{align}
g^{\{\Delta_{i}\}}_{\tau,\ell}(v,u)\approx v^{\frac{1}{2}\tau}\int_{0}^{\infty}\mathrm{ds}\frac{\sqrt{\ell} 2^{2 \ell+\tau -1} e^{-\frac{s^2+y}{s}} (\frac{1}{\ell})^{-a-b} (\frac{y}{s})^{-a-b}}{\sqrt{\pi } s}
=v^{\frac{1}{2}\tau}\frac{\sqrt{\ell} 2^{2 \ell+\tau } (\frac{1}{\ell})^{-a-b} y^{\frac{1}{2} (-a-b)} K_{a+b}(2 \sqrt{y})}{\sqrt{\pi }}.
\end{align}
Plugging in our values for $a$, $b$, and $y$ yields our final expression for crossed channel blocks in the $\ell\rightarrow\infty$ limit with $u\ell^{2}\lesssim \mathcal{O}(1)$:
\bea
g^{\{\Delta_{i}\}}_{\tau,\ell}(v,u)\approx v^{\frac{1}{2}\tau}\frac{\sqrt{\ell} 2^{2 \ell+\tau } u^{\frac{1}{4} (\Delta_{1}+\Delta_{2}-\Delta_{3}-\Delta_{4})} K_{\frac{1}{2} (-\Delta_{1}-\Delta_{2}+\Delta_{3}+\Delta_{4})}(2 \ell \sqrt{u})}{\sqrt{\pi }}.
\eea
This approximation breaks down when $u\ell^2 \gg 1$, but all of our sums are dominated by regions of fixed $u\ell^{2}$.

\section{Singularities in Direct and Crossed Channel}
\label{sec:Appendix_B}
A key result of this work is that to reproduce the identity block in the s-channel an infinite number of double-twist states are required in the t-channel. 
In the case of four identical scalars $\<\f\f\f\f\>$, this can be explained by the fact that the identity block is power law divergent in $u$ while the t-channel blocks have a $\log(u)$ divergence (see (\ref{eqn:spinning}), which reduces to this case with k=1 and $\s_{2}=\s_{1}=\Delta_{\f}$). Thus, any finite sum of the t-channel blocks cannot reproduce the s-channel contribution.

In the spinning case, the t-channel spinning conformal blocks are obtained by acting derivatives on the scalar blocks, which produces power law singularities in $u$ that can potentially become comparable to the s-channel divergences. 
If this were the case, then there may exist a solution to the crossing equation with a finite number of t-channel blocks. 
In this Appendix we will rule out this possibility.

First we need to look at the small u limit of the collinear t-channel conformal block, given by
\begin{align}
&v^{\frac{\tau}{2}} (1-u)^\ell  \, _2F_1\bigg(\frac{1}{2} (\tau + 2 \ell)+\frac{\Delta_{4}-\Delta_{1}}{2},\frac{1}{2} (\tau + 2 \ell)+\frac{\Delta_{3}-\Delta_{2}}{2};\tau + 2 \ell;1-u\bigg) 
\nonumber \\ &
\qquad \approx \pi  v^{\tau/2} \Gamma (2\ell+\tau) \csc \left(\frac{1}{2} \pi  (\Delta_{1}+\Delta_{2}-\Delta_{3}-\Delta_{4})\right) \times
\nonumber\\&
\qquad\bigg[\frac{1}{\Gamma (\frac{1}{2} (-\Delta_{1}-\Delta_{2}+\Delta_{3}+\Delta_{4}+2)) \Gamma (\frac{1}{2} (\Delta_{1}-\Delta_{4}+2\ell+\tau)) \Gamma (\frac{1}{2} (\Delta_{2}-\Delta_{3}+2\ell+\tau))} \nonumber\\&
\qquad -\frac{u^{\frac{1}{2} (\Delta_{1}+\Delta_{2}-\Delta_{3}-\Delta_{4})}}{\Gamma (\frac{1}{2} (\Delta_{1}+\Delta_{2}-\Delta_{3}-\Delta_{4}+2)) \Gamma (\frac{1}{2} (-\Delta_{1}+\Delta_{4}+2\ell+\tau)) \Gamma (\frac{1}{2} (-\Delta_{2}+\Delta_{3}+2\ell+\tau))}\bigg].
\end{align}
Clearly only the second term can lead to a singular behavior in a single spinning conformal block that matches the identity contribution. In the following calculations this will be the only term kept. 

We start with $\<JJ\f\f\>$ and the parity even double-twist states. Looking at the $P_{12} (Z_{1}\cdot Z_{2})$ structure in the even channel we get a contribution of order $u^{2-\Delta_{\f}}$, which is less singular then the $u^{-\Delta_{\f}}$ identity contribution. Similarly, the contribution to $(Z_{1}\cdot P_{2})(Z_{2}\cdot P_{1})$ is of order $u^{3-\Delta_{\f}}$. For the parity odd blocks the contribution to the $P_{12}(Z_{1}\cdot Z_{2})$ structure starts at order $u^{-\Delta_{\f}+2}$, while the contribution to $(Z_{1}\cdot P_{2})(Z_{2}\cdot P_{1})$ starts at order $u^{-\Delta_{\f}+3}$. Since we have to match all the dot products appearing in the identity piece, we only need to look at one structure and see that it is subleading for all the double-twist states to conclude that we cannot match the identity contribution with a finite number of blocks. 

For $\<TT\f\f\>$, the even and odd double-twist states contribute to $(Z_{1}\cdot P_{2})^2 (Z_{2}\cdot P_{1})^{2}$ starting at order $u^{5-\Delta_{\f}}$. The identity contribution has a power law singularity of order $u^{-\Delta_{\f}}$ in comparison, so we cannot match this with a finite number of spinning blocks.

Finally, we need to look at $\<JJJJ\>$. For simplicity we restrict to the $\U(1)$ case, but the symmetry group will not affect our results. Furthermore, we will need to be more careful with our approximation of the collinear block due to logarithmic singularities that can arise for special values of the dimensions, e.g. if they are all equal. To take into account these singularities we start with the hypergeometric form of the collinear blocks and do not expand in $u$ until after we act with the derivatives. The result is that both the twist 2 and 4 double-twist states contribute to the $(Z_{1}\cdot P_{2})(Z_{2}\cdot P_{1})(Z_{3}\cdot P_{4})(Z_{4}\cdot P_{3})$ structure at order $\log(u)$. The contribution of the twist 3 parity odd double-twist states to this structure vanishes at order $u^{-1}$, so we cannot match the $u^{-3}$ power law singularity from the identity channel. 

To conclude, no t-channel block is singular enough at small $u$ to match the identity contribution. Therefore we need an infinite number of states, which as we showed in the body of the text, has the spectrum of the double-twist states.

\section{Degeneracy Equations}\label{App:DegeneracyEquations}
Here we will review the degeneracy equations that appear in 3- and 4-point functions. 

We start by deriving the degeneracy relations among tensor structures appearing in the 3-point functions of spinning operators in three dimensions. These degeneracies arise because the 3-point structures depend on 6 vectors $\{ P_{i},Z_{i}\}$, which cannot be linearly independent in the 5-dimensional embedding space. The relation between the structures are found to be~\cite{Costa:2011dw,Costa:2011mg}:
\bea
(V_{1}H_{23}+V_{2}H_{13}+V_{3}H_{12}+2V_{1}V_{2}V_{3})^{2} = -2H_{12}H_{13}H_{23} + O(\{Z_{i}^{2},Z_{i}\cdot P_{i}\}).
\label{eqn:3D-deg2}
\eea
To prove this we embed the vectors in a 6d space so that they lie on the $x_{6}=0$ surface. It follows that $\epsilon(Z_{1},Z_{2},Z_{3},P_{1},P_{2},P_{3})=0$, or that the contraction of the vectors with the 6d epsilon tensor vanishes. Squaring this expression and using the identity 
\be\label{eqn:EpsilonEpsilon}
\epsilon(Z_{1},Z_{2},Z_{3},Z_{4},Z_{5},Z_{6})\epsilon(W_{1},W_{2},W_{3},W_{4},W_{5},W_{6})=\text{det}_{1\leq i,j\leq6}(Z_{i}\cdot W_{j}),
\ee
we obtain the above degeneracy. We will use this identity repeatedly to derive degeneracy conditions for the 4-point functions of spinning operators. 

Conformal invariance required that the 4-point tensor structures have the following properties: 
\bea
Q^{(k)}_{\chi_{1}\chi_{2}\chi_{3}\chi_{4}}(\{\lambda_{i}P_{i};\alpha_{i}Z_{i}\})=Q^{(k)}_{\chi_{1}\chi_{2}\chi_{3}\chi_{4}}(\{P_{i};Z_{i}\})\prod_{I}(\lambda_{i}\alpha_{i})^{\ell_{i}}.
\eea
The $Q^{(k)}(u,v)$ structures will be polynomials in the $H_{ij}$ and $V_{i,jk}$ tensor structures. There are additional degeneracy equations for the four point function. The first is that in general dimensions there are two independent $V_{i,jk}$ for each $i$. For example when $i=1$ we have
\bea
P_{23}P_{14}V_{1,23}+P_{24}P_{13}V_{1,42}+P_{34}P_{12}V_{1,34}=0, \label{eqn:deg}
\eea 
\noindent with related identities for $i=2,3,4$ related by permutation. Note that these conditions do not depend on the spacetime dimension. For $d<6$ there are more degeneracies among the tensor structures. The four point function depends on 8 vectors, the four pairs of position and polarization vectors, while the embedding space, if $d<6$, is at most 7-dimensional. We will only focus on $d=3$ here with the embedding space being 5d. We will use (\ref{eqn:EpsilonEpsilon}) with different vectors to derive the 4-point degeneracies.

For $\<JJ\f\f\>$ there are no linear relations among the tensor structures. This is easy to see, because the only nontrivial contraction of the vectors with the 6d epsilon tensor is $\epsilon(P_{1},P_{2},P_{3},P_{4},Z_{1},Z_{2})$, which must vanish. The only degeneracy conditions apart from (\ref{eqn:deg}) is then found from $\epsilon(P_{1},P_{2},P_{3},P_{4},Z_{1},Z_{2})^{2}=0$, and rewriting it in terms of dot products. This constraint is quadratic in $Z_{1}$ and $Z_{2}$, while the four point function $\<JJ\f\f\>$ is linear in both. Therefore, there are no degeneracies among the relevant tensor structures.

We now will consider possible degeneracies for the four point function tensor structures in $\<JJJJ\>$. The basic structures are
\bea
\{V_{1,23},V_{1,24},V_{2,34},V_{2,31},V_{3,41},V_{3,42},V_{4,12},V_{4,13},H_{12},H_{13},H_{14},H_{23},H_{24},H_{34}\},
\eea
\noindent out of which one can construct 43 distinct structures.

There are three degeneracy equations, linear in each $Z_{i}$, which follow from the fact that we have six 5d vectors which cannot be linearly independent:
\bea
\label{eq:firstdeg}\epsilon(P_{1},P_{2},P_{3},P_{4},Z_{1},Z_{2})\epsilon(P_{1},P_{2},P_{3},P_{4},Z_{3},Z_{4})=0, \\
\label{eq:seconddeg}\epsilon(P_{1},P_{2},P_{3},P_{4},Z_{1},Z_{3})\epsilon(P_{1},P_{2},P_{3},P_{4},Z_{2},Z_{4})=0, \\
\label{eq:thirddeg}\epsilon(P_{1},P_{2},P_{3},P_{4},Z_{1},Z_{4})\epsilon(P_{1},P_{2},P_{3},P_{4},Z_{2},Z_{3})=0.
\eea
Each individual contraction with the epsilon tensor vanishes and the product of two yields the degeneracy equations for the $H$ and $V$ structures. The three equations are not linearly independent; solving two implies the third. We will choose to solve for the latter two. Converting to the standard basis yields:
\begin{small}
\begin{align}\label{eq:secondlong}
&H_{12} v (H_{34} (-2 (u+1) v+(u-1)^2+v^2)+2 u V_{4,12} (V_{3,41} (u-v-1)+2 V_{3,42})-2 V_{4,13} (V_{3,41} (u+v-1)\nonumber \\& +V_{3,42} (u-v+1)))+H_{14} u (-H_{23} (u^2-2 u (v+1)+(v-1)^2)+2 v V_{3,41} (V_{2,34} (u-v+1)-2 V_{2,31})\nonumber \\& +2 V_{3,42} (V_{2,31} (-u+v+1)+V_{2,34} (u+v-1)))+2 (u V_{4,12} (H_{23} v V_{1,23} (u-v+1)+H_{23} V_{1,24} (u+v-1)\nonumber \\& -2 v V_{3,41} (V_{2,31} (V_{1,23}+V_{1,24})-V_{2,34} (u V_{1,23}-v V_{1,23}+V_{1,24}))+2 V_{3,42} (v V_{1,23} (V_{2,31}+V_{2,34})\nonumber \\& +V_{1,24} (V_{2,31}-V_{2,34})))+V_{4,13} (H_{23} u (-u V_{1,24}+v (V_{1,24}-2 V_{1,23})+V_{1,24})\nonumber \\& +2 v V_{3,41} (V_{1,23}-V_{1,24}) (V_{2,31}-u V_{2,34})-2 V_{3,42} (v (u V_{1,23} V_{2,34}+V_{1,23} V_{2,31}-V_{1,24} V_{2,31})\nonumber \\& +u V_{1,24} (V_{2,31}-V_{2,34})))+H_{34} v (V_{1,23} V_{2,31} (-(u+v-1))+u V_{1,23} V_{2,34} (u-v-1)\nonumber \\& +V_{1,24} V_{2,31} (-u+v-1)+2 u V_{1,24} V_{2,34}))=0,
\end{align}
\end{small}
\vspace{-.5cm}
\begin{small}
\begin{align}\label{eq:thirdlong}
&H_{12} v (H_{34} (u^2-2 u (v+1)+(v-1)^2)  +2 u V_{4,12} (V_{3,41} (u-v-1) + 2 V_{3,42}) - 2 V_{4,13} (V_{3,41} (u+v-1) \nonumber \\& +V_{3,42} (u-v+1)))+H_{13} u (-H_{24} v (u^2-2 u (v+1)+(v-1)^2)+2 u V_{4,12} (V_{2,34} (-(2 u+1) v+(u-1) u+v^2)\nonumber \\&+V_{2,31} (-u+v+1))+2 V_{4,13} (V_{2,31} (u+v-1)+u V_{2,34} (-u+v+1)))+\nonumber \\&2 (H_{24} v (u^2 v V_{1,23} V_{3,41}+u v V_{3,41} (-2 v V_{1,23}-V_{1,23}+V_{1,24})+u V_{3,42} (v V_{1,23}+V_{1,24})\nonumber \\&+(v-1) (v V_{1,23}-V_{1,24}) (v V_{3,41}-V_{3,42}))+H_{34} v (V_{1,23} V_{2,31} (-(u+v-1))+u V_{1,23} V_{2,34} (u-v-1)\nonumber \\&+V_{1,24} V_{2,31} (-u+v-1)+2 u V_{1,24} V_{2,34})-2 (u^3 v V_{1,23} V_{2,34} V_{3,41} V_{4,12}\nonumber \\&+u^2 (-2 v^2 V_{1,23} V_{2,34} V_{3,41} V_{4,12}-v (V_{4,12} (V_{1,23} V_{3,41} (V_{2,31}+V_{2,34})\nonumber \\&-V_{1,23} V_{2,34} V_{3,42}-V_{1,24} V_{2,34} V_{3,41})+V_{1,23} V_{2,34} V_{3,41} V_{4,13})+V_{1,24} V_{2,34} V_{3,42} V_{4,12})+\nonumber \\&u V_{4,12} (v V_{3,41} (v V_{2,34} (v V_{1,23}-V_{1,24})+(v+1) V_{1,23} V_{2,31})-V_{3,42} (V_{2,31} (v V_{1,23}+V_{1,24})\nonumber \\&+v V_{2,34} (v V_{1,23}-V_{1,24})))+u V_{4,13} (v V_{3,41} (V_{1,23} (v V_{2,34}+V_{2,31}+V_{2,34})-V_{1,24} V_{2,34})-V_{1,24} V_{2,34} V_{3,42})\nonumber \\&+(v-1) V_{2,31} V_{4,13} (v V_{1,23} V_{3,41}-V_{1,24} V_{3,42})))=0.
\end{align}
\end{small}
Using (\ref{eq:secondlong}) we can solve for $V_{1,23}V_{2,34}V_{3,41}V_{4,12}$ and using (\ref{eq:thirdlong}) we can solve for $H_{13}V_{2,34}V_{4,12}$. 
The reason for choosing these structures is as follows. For each equation we would like to solve for the tensor structure that will yield the most singular contribution to $H_{12}H_{34}$. That is, we want to take into account the behavior of the tensor structures themselves in the lightcone limit when solving the degeneracy equations. In practice, we then solve the above equations in terms of the dot products $(Z_{1}\cdot P_{3})(Z_{3}\cdot P_{1})(Z_{2}\cdot P_{4})(Z_{4}\cdot P_{2})$ and $(Z_{1}\cdot Z_{3})(Z_{2}\cdot P_{3})(Z_{4}\cdot P_{2})$ in the respective equations. Solving (\ref{eq:secondlong}) for  $(Z_{1}\cdot P_{3})(Z_{3}\cdot P_{1})(Z_{2}\cdot P_{4})(Z_{4}\cdot P_{2})$  will affect the large spin cross channel results, but not the s-channel. Solving  (\ref{eq:thirdlong}) for  $(Z_{1}\cdot Z_{3})(Z_{2}\cdot P_{3})(Z_{4}\cdot P_{2})$ will not affect either channel in the lightcone limit.

For $\<TT\f\f\>$, we choose the basis of structures to be $\{V_{1,23},V_{1,24},V_{2,31},V_{2,34},H_{12}\}$, from which one can construct the 14 four-point function structures:
\begin{align}
\{&H_{12}^2,H_{12} V_{1,24} V_{2,31},H_{12} V_{1,24} V_{2,34},V_{1,24}^2 V_{2,31}^2,V_{1,24}^2 V_{2,31} V_{2,34},V_{1,24}^2 V_{2,34}^2,H_{12} V_{1,23} V_{2,31},\nonumber \\
&H_{12} V_{1,23} V_{2,34},V_{1,23} V_{1,24} V_{2,31}^2,V_{1,23} V_{1,24} V_{2,31} V_{2,34},V_{1,23} V_{1,24} V_{2,34}^2,V_{1,23}^2 V_{2,31}^2,V_{1,23}^2 V_{2,31} V_{2,34},V_{1,23}^2 V_{2,34}^2\}.
\end{align}
There is a single degeneracy equation following from $\epsilon(P_{1},P_{2},P_{3},P_{4},Z_{1},Z_{2})^{2}=0$, which is
\begin{align}
&H_{12}^2 (u^2-2 u (v+1)+(v-1)^2)-4 H_{12} V_{2,31} (V_{1,23} (u+v-1)+V_{1,24} (u-v+1)) \nonumber
\\&+4 H_{12} u V_{2,34} (V_{1,23} (u-v-1)+2 V_{1,24})+4 (u V_{1,23} V_{2,34}-V_{1,23} V_{2,31}+V_{1,24} V_{2,31})^2=0.
\end{align}
Following the same logic as for $\<JJJJ\>$, we want to solve for the most singular tensor structure in the lightcone limit. Since the above equation must hold for all configurations and polarizations, we see that this structure must be $V_{1,23}^{2}V_{2,34}^{2}$. The degeneracy equation says $V_{1,23}^{2}V_{2,34}^{2}=-\frac{1}{4}u^{-2}H_{12}^{2}+ (...)$. Equivalently, we can expand the above equation in terms of the dot products and solve for $(Z_{1}\cdot P_{2})^2(Z_{2}\cdot P_{4})^{2}$ to find $(Z_{1}\cdot P_{2})^2(Z_{2}\cdot P_{4})^{2}=-\frac{1}{4u^{2}}(-2(P_{1}\cdot P_{2})(Z_{1}\cdot Z_{2}))^{2} + \mathcal{O}(u)$.

\section{3-point Functions and Differential Operators}
\label{App:3pfsDiffOps}
In this Appendix we provide more details about the structure of various 3-point functions in our analysis and the construction of the corresponding differential operators.

\subsection{$\<\f\f J\>$ and $\<\f\f T\>$}
For the scalars in the adjoint representation we have the general form
\bea
\<\f^{a}(P_{1})\f^{b}(P_{2})J^{c}(P_{3};Z_{3})\>=\hat{\lambda}_{\f\f J}f^{abc}\frac{V_{3}}{(P_{12})^{\Delta_{\f}-d/2}(P_{13})^{d/2}(P_{23})^{d/2}}.
\eea
The Ward identity implies $\hat{\lambda}_{\f\f J} = -\frac{1}{S_{d}}$. Given our normalization of the current, what appears in the conformal partial wave expansion is
\be
\lambda_{\f\f  J}=\frac{\hat{\lambda}_{\f\f J}}{\sqrt{C_{J}}}=-\frac{1}{S_{d}\sqrt{C_{J}}},
\ee
where $S_{d}$ gives the volume of $d-1$ dimensional sphere, \begin{math} S_{d}=\frac{2\pi^{\frac{d}{2}}}{\Gamma(\frac{d}{2})}\end{math}. $C_J$ is the current central charge and describes the normalization of the current 2-point function,
\bea
\<J(P_{1};Z_{1})J(P_{2};Z_{2})\>=C_{J}\frac{H_{12}}{(P_{12})^{d}}.
\eea

Similarly, for 3-point functions between scalars and the stress tensor we have
\bea
\< \f(P_{1})\f(P_{2})T(P_{3};Z_{3})\>=\hat{\lambda}_{\f\f T}\frac{V_{3}^{2}}{(P_{12})^{\Delta_{\f}-1-\frac{d}{2}}(P_{13})^{\frac{d}{2}+1}(P_{23})^{\frac{d}{2}+1}}, \nonumber 
\\
\hat{\lambda}_{\f\f T}= -\frac{\Delta_{\f}d}{(d-1)S_{d}},
\eea
where we use the normalization
\bea
\<T(P_{1};Z_{1})T(P_{2};Z_{2})\>=C_{T}\frac{H_{12}^{2}}{(P_{12})^{d+2}},
\eea
where $C_T$ is the central charge. The term appearing in the conformal partial wave expansion has an extra division by $\sqrt{C_{T}}$,
\bea
\lambda_{\f\f T}= -\frac{\Delta_{\f}d}{(d-1)S_{d}}\frac{1}{\sqrt{C_{T}}}. \qquad
\eea

\subsection{$\<JJJ\>$ and $\<JJT\>$}
We now present the differential representation of the parity preserving three point functions for $\<JJJ\>$ and $\<JJT\>$.

The parity preserving 3-point function for $\<J^{a}J^{b}J^{c}\>$ in embedding space is
\begin{eqnarray}
\<J^{a}(P_{1},Z_{1})J^{b}(P_{2},Z_{2})J^{c}(P_{3},Z_{3})\>=f^{abc}\frac{a_1V_{1}V_{2}V_{3} +a_2 H_{12}V_{3} +a_3 H_{13}V_{2} +a_4 H_{23}V_{1}}{(P_{12})^{\frac{d}{2}}(P_{13})^{\frac{d}{2}}(P_{23})^{\frac{d}{2}}},
\end{eqnarray}
\noindent where $f^{abc}$ are the structure constants. Conservation imposes $a_{2}=a_{3}=a_{4}$. The relation to the parametrization found in \cite{Osborn:1993cr}, Eq.~(3.10), is $a_{1}=a-2b$ and $a_{2}=-b$. The Ward identity further imposes that 
\begin{eqnarray}
S_{d}\bigg(\frac{1}{d}a+b\bigg)=C_{J},
\end{eqnarray}

We have labelled the OPE coefficient $b$ as $\lambda_{JJJ}$. The correct differential operator that reproduces (\ref{eqn:JJJ}) when acting on a scalar-scalar-current 3-point function is:  
\begin{eqnarray}
D_{L,J}=\frac{dC_{J}}{(d-2)S_{d}}(D_{12}D_{22}\Sigma_{L}^{0,2}+D_{11}D_{21}\Sigma_{L}^{2,0}-D_{12}D_{21}\Sigma_{L}^{1,1})+\frac{4\lambda_{JJJ}S_{d}-dC_{J}}{S_{d}(d-2)}D_{11}D_{22}\Sigma_{L}^{1,1}.\nonumber\\
\end{eqnarray}

We now proceed to study $\<JJT\>$. Without loss of generality we can now restrict to the case where $J$ is a $\U(1)$ current. Conformal invariance and symmetry under $1\leftrightarrow2$ implies
\bea
\<J(P_{1};Z_{1})J(P_{2};Z_{2})T(P_{3};Z_{3})\>=\frac{\alpha V_{1}V_{2}V_{3}^{2}+\beta(H_{13}V_{2}+H_{23}V_{1})V_{3}+\gamma H_{12}V_{3}^{2}+\eta H_{13}H_{23}}{(P_{12})^{\frac{d}{2}-1}(P_{13})^{\frac{d}{2}+1}(P_{23})^{\frac{d}{2}+1}} .\nonumber
\\
\eea
Conservation implies 
\bea
-\alpha-d\beta+(2+d)\gamma=0, \nonumber \\
-2\beta+2\gamma+(2-d)\eta=0.
\eea
The implications of conservation for $\<JJT\>$ was first solved in \cite{Osborn:1993cr}, see Eqs.~(3.11-3.14). The relations between our parametrization and theirs is given by
\bea
\begin{split}
&\eta=2e, \ \ \ \beta=-2c,  \\
&\gamma= a-\frac{b}{d}-\frac{4c}{d}, \ \ \   \alpha=2a+b\big(1-\frac{2}{d}\big)-\frac{8c}{d},
\end{split}
\eea
\noindent where the parameters $a$ and $c$ used to parametrize $\<JJT\>$ are unrelated to those used in $\<JJJ\>$. Furthermore they solved the Ward identities to find
\bea
2S_{d}(c+e)=dC_{J}.
\eea
\noindent So $\<JJT\>$ is fixed up to one OPE coefficient and $C_{J}$. We labelled the OPE coefficient $c$ as $\lambda_{JJT}$ in the body of the paper. The required differential operator is then found to be 
\bea
D_{L,T}=\bigg[\bigg(2\lambda_{JJT}-\frac{C_{J}d(d-2)}{(d-1)S_{d}}\bigg)D_{11}D_{22}+\bigg(2\lambda_{JJT}+\frac{C_{J}d^{2}}{S_{d}(1-d)}\bigg)D_{12}D_{21}-2\lambda_{JJT}H_{12}\bigg]\Sigma_{L}^{1,1}. \hspace{0.5cm}
\eea

\subsection{Differential Operators for $\<TTT\>$ and $\<T\f[T\f]\>$}\label{App:TTffDifferentialOperators}
We will start by analyzing $\<TTT\>$ in the standard basis and then the differential basis. Restricting to parity-preserving correlators, the allowed tensor structures are
\begin{align}
&Q_{1}=V_{1}^{2}V_{2}^{2}V_{3}^{2}, \label{eqn:stand1} \\
&Q_{2}=H_{23}V_{1}^{2}V_{2}V_{3}+H_{13}V_{1}V_{2}^{2}V_{3},\\
&Q_{3}=H_{12}V_{1}V_{2}V_{3}^{2},\\
&Q_{4}=H_{12}H_{13}V_{2}V_{3}+H_{12}H_{23}V_{1}V_{3},\\
&Q_{5}=H_{13}H_{23}V_{1}V_{2},\\
&Q_{6}=H_{12}^{2}V_{3}^{2},\\
&Q_{7}=H_{13}^{2}V_{2}^{2}+H_{23}^{2}V_{1}^{2},\\
&Q_{8}=H_{12}H_{13}H_{23}. \label{eqn:stand8}
\end{align}
The $H_{12}H_{13}H_{23}$ structure is not linearly independent in three dimensions, as follows from Eq.~(\ref{eqn:3D-deg}). Above we only required symmetry under interchange between $1\leftrightarrow2$. We could have also required symmetry under $2\leftrightarrow3$, but the above basis is simpler when comparing the results to \cite{Osborn:1993cr} where the latter symmetry was obscured. 

The constraints of conservation were solved in \cite{Osborn:1993cr} for general dimensions where they parametrized the correlation function in terms of 8 variables: $a$, $b$, $b'$, $c$, $c'$, $e$, $e'$, and $f$. These parameters are unrelated to those appearing in the $\<JJJ\>$ and $\<JJT\>$ correlation functions. Labelling the coefficients of $Q_{i}$ by $x_{i}$, the mapping between the bases is given by
\bea
x_{1}=8(c+e)+f, \ \ \
x_{2}=-4(4b'+e'),\ \ \
x_{3}=4(2c+e),\\
x_{4}=-8b', \qquad \qquad
x_{5}=8b+16a, \qquad \qquad \ \ 
x_{6}=2c,\\
x_{7}=2c', \qquad \qquad \qquad
x_{8}=8a.
\eea
Conservation at $P_{1}$ or $P_{2}$ imposes that
\bea
x_{1}=2x_{2}+\frac{1}{4}(d^{2}+2d-8)x_{4}-\frac{1}{2}d(2+d)x_{7}, \qquad
x_{8}=\frac{x_{2}-(\frac{d}{2}+1)x_{4}+2dx_{7}}{\frac{d^{2}}{2}-2},\\
x_{2}=x_{3},\qquad 
x_{4}=x_{5},\qquad 
x_{6}=x_{7},
\eea
which is consistent with the conservation conditions found in \cite{Osborn:1993cr}.

Furthermore, they found that the Ward identity constraints are given by 
\bea
4S_{d}\frac{(d-2)(d+3)a-2b-(d+1)c}{d(d+2)}=C_{T}.
\eea
Imposing the Ward identity and using the degeneracy equation (\ref{eqn:3D-deg}), we find in $d=3$
\begin{align}
&x_{1}=147\lambda_{TTT}-\frac{405C_{T}}{16\pi},\\
&x_{2}=x_{3}=52\lambda_{TTT}-\frac{75C_{T}}{8\pi},\\
&x_{4}=x_{5}=16\lambda_{TTT}-\frac{15C_{T}}{4\pi}, \\
&x_{6}=x_{7}=-2\lambda_{TTT}, \\
&x_{8}=0,
\end{align}
\noindent where $\lambda_{TTT}=2a-c$. As expected, we find that in three dimensions the parity-even part of $\<TTT\>$ has two linearly independent forms, which we parametrize with $\lambda_{TTT}$ and $C_{T}$. Note that in~\cite{Buchel:2009sk} the extra parameter was called $t_4$. Our parametrization is related to theirs by the relation $\lambda_{TTT} = \frac{3 C_T (60-t_4)}{2^{10}\pi}$.

We now need to find the mapping between the standard basis and the differential basis. An over-complete differential basis symmetric under $1\leftrightarrow2$ is given by:
\begin{align}
&W_{1}=D_{11}^{2}D_{22}^{2}\Sigma_{L}^{2,2},\\ \label{eqn:Diff1}
&W_{2}=H_{12}D_{11}D_{22}\Sigma_{L}^{2,2},\\
&W_{3}=D_{21}D_{11}^{2}D_{22}\Sigma_{L}^{3,1}+D_{12}D_{22}^{2}D_{11}\Sigma_{L}^{1,3} ,\\
&W_{4}=H_{12}(D_{21}D_{11}\Sigma_{L}^{3,1}+D_{12}D_{22}\Sigma_{L}^{1,3}),\\
&W_{5}=D_{12}D_{21}D_{11}D_{22}\Sigma_{L}^{2,2},\\
&W_{6}=H_{12}^{2}\Sigma_{L}^{2,2},\\
&W_{7}=D_{21}^{2}D_{11}^{2}\Sigma_{L}^{4,0}+D_{12}^{2}D_{22}^{2}\Sigma_{L}^{0,4},\\
&W_{8}=H_{12}D_{12}D_{21}\Sigma_{L}^{2,2},\\ \label{eqn:Diff8}
&W_{9}=D_{12}^{2}D_{21}^2\Sigma_{L}^{2,2},\\
&W_{10}=D_{12}D_{21}^{2}D_{11}\Sigma_{L}^{3,1}+D_{21}D_{12}^{2}D_{22}\Sigma_{L}^{1,3}.
\end{align}
Although there are 10 possible differential operators, only the first 8 are required to express $\<TTT\>$ in terms of differential operators acting on a scalar structure. That is, we can find an invertible matrix such that
\bea
W_{i}=\sum_{j=1}^{8}a_{ij}Q_{j} \quad \text{and} \quad Q_{i}=\sum_{j=1}^{8}(a^{-1})_{ij}W_{j}.
\eea
The matrix $(a^{-1})_{ij}$ in general dimensions is given in Appendix \ref{sec:Appendix_C}. The differential representation of $\<TTT\>$ is then given by 
\bea
\mathcal{D}_{T}=\sum_{ij}x_{i}(a^{-1})_{ij}W_{j}.
\eea
Taking into account our normalization for $\<TT\>$, what appears in the conformal block expansion is then $\frac{1}{\sqrt{C_{T}}}\mathcal{D}_{T}$.

Let us consider the 3-point functions $\<T\f\mathcal{O}^{(\ell)}\>$ where $\mathcal{O}^{(\ell)}$ are double-twist states. The two possible operators take the schematic form $[T\f]_{n,\ell}=T_{\mu\nu}(\partial^{2})^{n}\partial_{\sigma_{1}}...\partial_{\sigma_{\ell-2}}\f$ and $\widetilde{[T\f]}_{n,\ell}=\epsilon_{\alpha}^{\ \nu\kappa}T_{\mu\nu}(\partial^{2})^{n}\partial_{\kappa}\partial_{\sigma_{1}}...\partial_{\sigma_{\ell-2}}\f$, with twists 2+2$n$ and 3+2$n$ respectively. Below we will restrict to the $n=0$ operators.

Starting with the parity even differential operators, the most general operator is 
\bea
f_{1}D_{11}^{2}\Sigma_{L}^{2,0} + f_{2}D_{12}D_{11}\Sigma_{L}^{1,1} +f_{3} D_{12}^{2}\Sigma_{L}^{0,2}.
\eea
Imposing conservation yields
\bea
f_{1}=\frac{6f_{3}}{(\Delta_{\f}+\ell-2)(\Delta_{\f}+\ell-1)}, \qquad f_{2}=-\frac{4f_{3}}{\Delta_{\f}+\ell-1}. 
\eea
For the parity odd states, the differential operator has the form
\bea
t_{1}D_{11}\tilde{D}_{1}\Sigma_{L}^{1,0}+t_{2}D_{12}\tilde{D}_{1}\Sigma_{L}^{0,1}.
\eea
Conservation implies
\bea
t_{1}=-\frac{3t_{2}}{\Delta_{\f}+\ell-1}.
\eea
The t-channel left differential operators are constructed in the usual way by letting $2\leftrightarrow4$ in the definition of the differential building blocks. The right differential operators are then constructed from the left operators by letting $1\rightarrow2$ and $3\rightarrow4$. The end result is
\begin{align}
D^t_{[T\f]_{0,\ell}}=&\bigg(\frac{6}{(\Delta_{\f}+\ell-2)(\Delta_{\f}+\ell-1)}(D^{t}_{11})^{2}\Sigma_{L}^{2,0}-\frac{4}{\Delta_{\f}+\ell-1} D^{t}_{14}D^{t}_{11}\Sigma_{L}^{1,1}+(D^{t}_{14})^{2}\Sigma_{L}^{0,2}\bigg)\nonumber
\\&\bigg(\frac{6}{(\Delta_{\f}+\ell-2)(\Delta_{\f}+\ell-1)}(D^{t}_{22})^{2}\Sigma_{R}^{2,0}-\frac{4}{\Delta_{\f}+\ell-1} D^{t}_{23}D^{t}_{22}\Sigma_{L}^{1,1}+(D^{t}_{23})^{2}\Sigma_{R}^{0,2} \bigg),
\\
D^t_{\widetilde{[T\f]}_{0,\ell}}=&\bigg(-\frac{3}{\Delta_{\f}+\ell-1}D^{t}_{11}\tilde{D}^{t}_{1}\Sigma_{L}^{1,0}+D^{t}_{14}\tilde{D}^{t}_{1}\Sigma_{L}^{0,1}\bigg)\bigg(-\frac{3}{\Delta_{\f}+\ell-1}D^{t}_{22}\tilde{D}^{t}_{1}\Sigma_{R}^{1,0}+D^{t}_{23}\tilde{D}^{t}_{1}\Sigma_{R}^{0,1}\bigg).
\end{align}

\section{Charge 1-point Function}\label{App:Charge1PF}
In this Appendix, we compute the charge flux 1-point function in general dimensions. The charge flux 1-point function was defined in \cite{Hofman:2008ar}. In a CFT with a non-Abelian global symmetry $G$, we inject a unit amount of charge with a local perturbation $\epsilon^i J_{i}^{+}$ at the origin, where the ``+" indicates that the operator carries charge $+1$ under a chosen $\U(1)\subset G$. The perturbation propagates and carries the charge away to infinity. A charge detector at spatial infinity along the direction of a unit vector $\vec{n}$ will detect an integrated charge flux given by
\begin{eqnarray}
\langle\mathcal{Q}(n)\rangle=\frac{1}{S_{d}}\bigg(1+\tilde{a}_{2}\bigg(\frac{|\epsilon\cdot n|^{2}}{|\epsilon|^{2}}-\frac{1}{d-1}\bigg)\bigg).
\end{eqnarray}
The second piece integrates to zero and characterizes the asymmetry in the charge flux. $\tilde{a}_2$ is a coefficient determined by the microscopic theory. In this Appendix, we determine $\tilde{a}_2$ in a free theory involving $N_{df}$ Dirac fermions and $N_s$ scalars transforming under the global symmetry. These numbers are counted using the index of the representations~\cite{Osborn:1993cr},
\be
\Tr(t_{s}^{a} t_{s}^{b}) = N_{s} \delta^{ab}, \hspace{1cm} \Tr(t_{df}^{a} t_{df}^{b}) = N_{df} \delta^{ab}.
\ee
Our method is an extension of the Appendix C of \cite{Maldacena:2011jn} to the case of spin-1 currents.

In free theories, $\tilde{a}_{2}$ has the following general form:
\begin{eqnarray}
\tilde{a}_{2}=\frac{c_{1}N_{s}+c_{2}N_{df}}{C_{J}}.
\end{eqnarray}
This follows from the definition of the charge correlator, since the three point function $\<JJJ\>$ in a free theory is linear in $N_{s}$ and $N_{df}$. The $C_{J}$ in the denominator comes from normalizing the charge correlator with the two point function $\<JJ\>$~\cite{Osborn:1993cr}:
\be
C_J=\frac{1}{S^{2}_{d}}\left(\frac{N_s}{d-2}+N_{df} 2^{\left\lfloor \frac{d}{2}\right\rfloor} \right)
\ee
Note that the $\gamma$ matrixes are $\left\lfloor \frac{d}{2}\right\rfloor \times \left\lfloor \frac{d}{2}\right\rfloor$ in $d$ dimensions. 
We argue that in a theory of free bosons it is impossible to create two particles propagating back to back perpendicular to the direction of the current, so that the charge correlator has to vanish at $n\cdot\epsilon=0$. 
We create a state with $J^{+}_{1}$ and consider the matrix element $\<p,-p|\epsilon \cdot J^+|0\>$, where $\<p,-p|$ denotes a two particle state with $p_1=0$. 
Under a reflection of the first axis, $J^{+}_{1}\rightarrow -J^{+}_{1}$ and $p \rightarrow p$. 
Therefore the matrix element vanishes because the operator is antisymmetric but the state is symmetric under this reflection. 
A similar argument indicates that the fermion charge 1-point function should vanish when $\epsilon$ and $n$ are parallel.

This fixes 
\be
c_{1}=\frac{1}{S^{2}_{d}}\frac{d-1}{d-2}, \hspace{1cm} c_{2}=-\frac{2^{\left\lfloor \frac{d}{2}\right\rfloor}}{S^{2}_{d}}\frac{d-1}{d-2},
\ee
so consequently we find
\begin{eqnarray}
\tilde{a}_{2}= (d-1)\frac{ N_{s}- N_{df} 2^{\left\lfloor \frac{d}{2}\right\rfloor }}{N_{s}+(d-2) N_{df} 2^{ \left\lfloor \frac{d}{2}\right\rfloor }} .
\end{eqnarray}

Setting $d=4$, we obtain $\tilde{a}_{2}=3-\frac{36 N_{df}}{8 N_{df}+N_{s}}$. This matches with the result in 4 dimensions given in~\cite{Hofman:2008ar}. 
We also see in general dimensions that if we have an equal number of on-shell bosonic and fermionic degrees of freedom in a representation, $\tilde{a}_2$ vanishes and the result is a uniform distribution. 

Setting $d=3$, we see that the charge is always positive if $\tilde{a}_{2}$ lies between the free field theory values in 3 dimensions. In a theory with only scalars charged under the relevant global symmetry, there is a zero at $\theta=\pi/2$ and if there are only charged fermions there is a zero at $\theta=0,\pi$. 

We can rewrite this in terms of the coefficients $C_J$ and $\lambda_{JJJ}$ that parameterize $\<JJJ\>$. 
In a free theory, we have: 
\begin{align}
\lambda_{JJJ}&=\frac{N_{s}}{2 (d-2) S_{d}^3}+\frac{N_{df} 2^{\left\lfloor{d/2}\right\rfloor}}{S_{d}^{3}},\\
C_{J}&=\frac{N_{s}}{(d-2) S_{d}^2}+\frac{N_{df} 2^{\left\lfloor{d/2}\right\rfloor}}{S_{d}^2}.
\end{align}
We then obtain
\begin{eqnarray}
\tilde{a}_{2}=\frac{d-1}{d-2}\left(2 d-3- \lambda_{JJJ}\frac{2 S_{d}}{C_J} (d-1)\right).
\end{eqnarray}
If we require $\langle\mathcal{Q}(\vec{n})\rangle$ to be non-negative for all $\vec{n}$, then 
\be
\frac{C_J}{2S_d} \le \lambda_{JJJ} \le \frac{C_J}{S_d}.
\ee
For $d=3$, this agrees with (\ref{eqn:boundsForJJJ}).

\section{Other Technical Details}
\label{sec:Appendix_C}
Here we will collect some other formulas referenced in the body of the text.

\subsection{Change of Basis for $\<TTT\>$}
In general dimensions the matrix $(a^{-1})_{ij}$ is given by
\begin{align}
\begin{tiny}
\left(
\begin{array}{cccccccc}
 -\frac{1}{-2 h^4-3 h^3+h} & \frac{1-h}{-2 h^4-3 h^3+h} & \frac{1}{-2 h^4-h^3+h^2} & \frac{1}{h (2 h^2+h-1)} & -\frac{2}{-2 h^4-3 h^3+h} & \frac{1}{2 h^2+2 h} & \frac{1}{4 h^4-6 h^3+2 h^2} & \frac{1}{h (2 h^2+h-1)} \\
 \frac{h+3}{-2 h^4-3 h^3+h} & \frac{h (h+2)+5}{h (h+1)^2 (2 h-1)} & \frac{2 (h+2)}{h^2 (2 h^2+h-1)} & \frac{2}{-2 h^3-h^2+h} & \frac{2 (h+3)}{-2 h^4-3 h^3+h} & -\frac{1}{h^2+h} & -\frac{h+4}{4 h^4-6 h^3+2 h^2} & \frac{h+3}{-2 h^3-h^2+h} \\
 0 & \frac{1}{2 h-4 h^2} & 0 & \frac{1}{2 h-4 h^2} & 0 & -\frac{1}{2 h} & 0 & \frac{1}{2 h-4 h^2} \\
 0 & \frac{1}{2 h-1}-\frac{1}{h} & 0 & -\frac{1}{h-2 h^2} & 0 & \frac{1}{h} & 0 & -\frac{h+1}{h-2 h^2} \\
 \frac{h (h+2)+3}{h (h+1)^2 (2 h-1)} & \frac{-h^3+5 h^2+9 h+11}{-4 h^4-6 h^3+2 h} & -\frac{h (h+3)+4}{h^2 (2 h^2+h-1)} & -\frac{(h-1) (h+3)}{2 h (2 h^2+h-1)} & \frac{h (h+2)+5}{h (h+1)^2 (2 h-1)} & \frac{1}{2 h}-\frac{1}{h+1} & \frac{h+2}{2 h^4-3 h^3+h^2} & \frac{h (h+2)+5}{2 h (2 h^2+h-1)} \\
 0 & 0 & 0 & 0 & 0 & 1 & 0 & 0 \\
 \frac{2 (h+3)}{h (2 h^2+h-1)} & -\frac{4 (h+3)}{h (2 h^2+h-1)} & \frac{2 h+8}{h^2-2 h^3} & 0 & \frac{8}{h (2 h^2+h-1)} & 0 & \frac{h^2+3 h+4}{2 h^4-3 h^3+h^2} & -\frac{4}{h-2 h^2} \\
 0 & \frac{(h-1)^2}{h-2 h^2} & 0 & \frac{1-h^2}{h-2 h^2} & 0 & \frac{h-1}{h} & 0 & \frac{h^2+1}{h-2 h^2} \\
\end{array}
\right)
\end{tiny},
\end{align}

where $h=\frac{d}{2}$ and the matrix maps the differential basis $W_{i}$ (Eqs.~(\ref{eqn:Diff1}-\ref{eqn:Diff8})) to the standard basis $Q_{i}$ (Eqs.~(\ref{eqn:stand1}-\ref{eqn:stand8})).

\subsection{$\SU(N)$ Adjoint Crossing Matrix}
The matrix $\mathcal{M}$ used in the crossing symmetry equation for four $\SU(N)$ adjoints is given by
\bea\label{eqn:SUNAdjCrossingEquations}
\mathcal{M}^{r'}_{r} &=&
\left(
\setlength\arraycolsep{1pt}\begin{array}{cccccc}
 \frac{1}{(N-1) (N+1)} & \frac{2 N}{(N-1) (N+1)} & \frac{2 (N-2) (N+2)}{(N-1) N (N+1)} & \frac{(N-2) (N+2)}{(N-1) (N+1)} & \frac{(N-3) N^2}{(N-1)^2 (N+1)} & \frac{N^2 (N+3)}{(N-1) (N+1)^2} \\
 \frac{1}{2 N} & \frac{1}{2} & \frac{(N-2) (N+2)}{2 N^2} & 0 & \frac{N-3}{2 (N-1)} & -\frac{N+3}{2 (N+1)} \\
 \frac{N}{2 (N-2) (N+2)} & \frac{N^2}{2 (N-2) (N+2)} & \frac{N^2-12}{2 (N-2) (N+2)} & -\frac{N}{(N-2) (N+2)} & -\frac{(N-3) N^3}{2 (N-2)^2 (N-1) (N+2)} & \frac{N^3 (N+3)}{2 (N-2) (N+1) (N+2)^2}
   \\
 \frac{1}{2} & 0 & -\frac{2}{N} & \frac{1}{2} & -\frac{(N-3) N}{2 (N-2) (N-1)} & -\frac{N (N+3)}{2 (N+1) (N+2)} \\
 \frac{1}{4} & \frac{1}{2} & -\frac{N+2}{2 N} & -\frac{N+2}{4 N} & \frac{N^2-N+2}{4 (N-2) (N-1)} & \frac{N+3}{4 (N+1)} \\
 \frac{1}{4} & -\frac{1}{2} & \frac{N-2}{2 N} & -\frac{N-2}{4 N} & \frac{N-3}{4 (N-1)} & \frac{N^2+N+2}{4 (N+1) (N+2)} \\
\end{array}
\right)\nn,\\
\eea

in the basis $r = \big(I\,, Adj_a\,, Adj_s\,, (S,\bar{A})_a \oplus (A,\bar{S})_a\,, (A,\bar{A})_s\,, (S,\bar{S})_s\big)$.

\bibliography{Biblio}{}
\bibliographystyle{utphys}

\end{document}